\documentclass[aps,prapplied,showpacs,twocolumn,preprintnumbers,amsmath,amssymb,longbibliography]{revtex4-2}

\usepackage{graphicx}
\usepackage{dcolumn}
\usepackage{bm}
\usepackage{color}
\usepackage{amsmath}
\usepackage{upgreek}
\usepackage{float}
\usepackage[colorlinks,linkcolor=blue, urlcolor=blue, anchorcolor=blue, citecolor=blue]{hyperref}

\begin{document}

\title{Experimental Observation of Phase Transitions in Spatial Photonic Ising Machine}

\author{Yisheng Fang}
\thanks{These authors contributed equally to this work.}
\affiliation{Interdisciplinary Center of Quantum Information, State Key Laboratory of Modern Optical Instrumentation, and Zhejiang Province Key Laboratory of Quantum Technology and Device, Department of Physics, Zhejiang University, Hangzhou 310027, China}

\author{Junyi Huang}
\thanks{These authors contributed equally to this work.}
\affiliation{Interdisciplinary Center of Quantum Information, State Key Laboratory of Modern Optical Instrumentation, and Zhejiang Province Key Laboratory of Quantum Technology and Device, Department of Physics, Zhejiang University, Hangzhou 310027, China}

\author{Zhichao Ruan}
\email{zhichao@zju.edu.cn}
\affiliation{Interdisciplinary Center of Quantum Information, State Key Laboratory of Modern Optical Instrumentation, and Zhejiang Province Key Laboratory of Quantum Technology and Device, Department of Physics, Zhejiang University, Hangzhou 310027, China}

\begin{abstract}
Statistical spin dynamics plays a key role to understand the working principle for novel optical Ising machines. Here we propose the gauge transformations for spatial photonic Ising machine, where a single spatial phase modulator simultaneously encodes spin configurations and programs interaction strengths. Thanks to gauge transformation, we experimentally evaluate the phase diagram of high-dimensional spin-glass equilibrium system with $100$ fully-connected spins. We observe the presence of paramagnetic, ferromagnetic as well as spin-glass phases and determine the critical temperature $T_c$ and the critical probability ${{p}_{c}}$ of phase transitions, which agree well with the mean-field theory predictions. Thus the approximation of the mean-field model is experimentally validated in the spatial photonic Ising machine. Furthermore, we discuss the phase transition in parallel with solving combinatorial optimization problems during the cooling process and identify that the spatial photonic Ising machine is robust with sufficient many-spin interactions, even when the system is associated with the optical aberrations and the measurement uncertainty.

\end{abstract}
\maketitle

As a promising approach to solve a large class of NP-hard problems \cite{lucas2014ising}, recently it has attracted tremendous interest to simulate spin glass Hamiltonians in unconventional computing architectures, including optical parametric oscillators \cite{mcmahon2016fully,inagaki2016coherent,inagaki2016large,bohm2019poor,marandi2014network}, lasers \cite{utsunomiya2011mapping, babaeian2019single, tradonsky2019rapid,parto2020realizing, honari2020mapping}, polariton \cite{kalinin2020polaritonic, berloff2017realizing, kalinin2018simulating},  trapped ions \cite{kim2010quantum}, atomic and photonic condensates \cite{struck2013engineering,kassenberg2020controllable}, electronic memorisers \cite{cai2020power}, superconducting qubits \cite{johnson2011quantum, boixo2014evidence, king2018observation}, and nanophotonics circuits \cite{roques2020heuristic,prabhu2020accelerating,shen2017deep,wu2014optical, okawachi2020demonstration,prabhu2020accelerating}. In particular, the spatial photonic Ising machine with optical modulation in spatial domain has been demonstrated with reliable large-scale Ising spin systems, even up to thousands of spins \cite{pierangeli2019large}. Like spatial analog computations \cite{silva2014performing,bykov2014optical,ruan2015spatial, youssefi2016analog,zhu2017plasmonic,zhang2018implementing,guo2018photonic,zhu2019generalized,zangeneh2020analogue}, the setup benefits from the high speed and parallelism of optical signal processing. Thus spatial photonic Ising machines demonstrate high efficiency in searching the ground states and therefore solving the combinatorial optimization problems \cite{pierangeli2020noise, pierangeli2020adiabatic,pierangeli2020scalable,kumar2020large}.

Generally, complete characterization of possible stable phases is necessary to estimating the Ising description for practical systems \cite{nishimori2001statistical}. Also statistical spin dynamics about phase transitions plays a key role to understand the working principle in  spin systems \cite{mertens1998phase, wang2013coherent, strinati2019theory}. However, it is challenging to explore all controlled parameters and diagram stable phases for proposed Ising machines from either theoretical or experimental perspective. In the theoretical way, conventionally, a mean-field model is required with the approximation of many-body interaction by one-body average, and such a hypothesis needs to be verified by experimental investigations \cite{nishimori2001statistical,leuzzi2009phase}. On the other hand, since there are enormous spin configurations when a system has  a large number of spins, experimental investigations for phase transition are typically limited for the systems with few spins. For example, the phase diagram was experimentally investigated for the simplest coherent Ising machine with two-spin coupled parametric oscillators \cite{bello2019persistent}.

In this Letter, we focus on spatial photonic Ising machine and investigate the phase transitions in spatial spin glass systems. Here we propose a gauge transformation to incorporate both the spin configuration and interaction strengths. Thanks to the gauge transformation, by performing the spin system in equilibrium states, we experimentally demonstrate the phase diagram with the spin number as large as $N=100$. We observe the presence of paramagnetic, ferromagnetic as well as spin-glass phases and determine the critical temperature $T_c$ and the critical probability ${{p}_{c}}$ of phase transitions, which agree well with the mean-field theory predictions. Thus we experimentally verify the approximation of the mean-field model in the spatial photonic Ising machine. Furthermore, we discuss the impact of the phase transition in parallel with solving combinatorial optimization problems. We identify strong fluctuations of Ising energy when the temperature is close to $T_c$, and thus the system needs more time to return to equilibrium after perturbations. Below $T_c$, the system can be dramatically cooled down to the ground state, which indicates that the spatial photonic Ising machine is expected to be robust, even when the system has the optical aberrations and the measurement uncertainty.

\textit{Gauge transformation.}---We first consider the spatial photonic Ising machine proposed in Ref.~\cite{pierangeli2019large, pierangeli2020noise}. The collimated laser beam with uniform and unitary amplitude illuminates an amplitude spatial modulator in order to generate the amplitude modulation $\left\{ {{\xi }_{j}} \right\}$ and $-1\le {{\xi }_{j}}\le 1$. Then through the pixel alignment, the modulated beam impinges on the phase-only spatial light modulator (SLM), where the spin configuration $\mathbf{S}\text{=}\left\{ {{\sigma }_{j}} \right\}$ is encoded on the beam wavefront through the phase modulation as ${{\varphi }_{j,\text{SLM}}}\text{=}{{\sigma }_{j}}\frac{\pi }{2}$. Here, ${{\sigma }_{j}}$ takes binary value of either $+1$ or $-1$, and ${{\varphi }_{j,\text{SLM}}}$ is the phase of the $j\text{th}$ pixel on SLM. Subsequently, a lens performs Fourier transformation of the optical field, then the normalized field intensity at the focal plane, $I(u,v)$, is detected by the CCD. Especially, the intensity ${{I}_{0}}$ at the center where $(u,v)=(0,0)$ is contributed by the interactions between every two spins as ${{I}_{0}}=\sum\limits_{ij}{{{\xi }_{i}}{{\xi }_{j}}{{\sigma }_{i}}{{\sigma }_{j}}}$ (see Supplementary Material Sec.~I for details). Thus the Hamiltonian of such a Mattis-type spin glass system can be defined as
\begin{equation}
H=-J I_0=-\sum\limits_{ij}{{{J}_{ij}}{{\sigma }_{i}}}{{\sigma }_{j}} \label{eq:1}
\end{equation}
where $J$ is a constant with the unit of energy, and ${{J}_{ij}}=J{{\xi }_{i}}{{\xi }_{j}}$ is the interaction strengths between the spins.

\begin{figure}
\centerline{\includegraphics[width=3.2in] {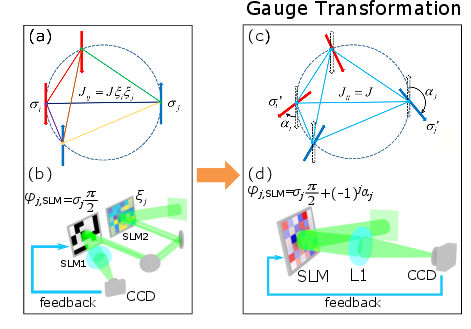}}
\caption{ \label{fig:1} Schematic of the gauge transformation and the experimental optical setups for spatial spin glass systems. (a,b) Without gauge transformation, the spin $\mathbf{S}\text{=}\left\{ {{\sigma }_{j}} \right\}$ and the interaction strengths {} are separately encoded on a phase spatial light modulator (SLM1) and an amplitude one (SLM2), whose the phase and amplitude modulation are ${{\varphi }_{j,\text{SLM}}}$ and ${{\xi }_{j}}$, respectively. (c, d) By gauge transformation, the gauge-transformed effective spin configuration ${{\mathbf{{S}'}}^{z}}=\left\{ {\sigma'_j}^{z} \right\}$ is encoded through only one phase SLM, following Eq. (\ref{eq:4}). The modulated light is detected by a CCD camera in the back-focus plane of a lens L1. For both two cases, the SLMs are illuminated by collimated laser beams. Details of the experimental setup are presented in SM Sec.~III.}
\end{figure}

Now let us propose the gauge transformation such that the inhomogeneous interaction strengths ${{J}_{ij}}$ can be transformed into the spin orientations. As shown in Fig.~\ref{fig:1}(c), each original spin ${{\sigma }_{j}}$ is rotated clockwise with respect to the $z$-axis with the angle ${{\alpha }_{j}}=\arccos {{\xi }_{j}}$ to arrive at a new spin vector $\bm{\sigma_j}'$, then $\bm{\sigma_j }'$ is projected on the $z$-axis to obtain the effective spin ${\sigma'_j}^{z}={{\xi }_{j}}{{\sigma }_{j}}$ and ${{\mathbf{{S}'}}^{z}}=\left\{ {\sigma'_j}^{z} \right\}$ is the gauge-transformed effective spin configuration. As the results, the interactions between the $z$ components of gauge-transformed spins become uniform in both short and long ranges, with the strength of $J$. The gauge transformation above is given as
\begin{equation}
\sigma_j \to {\sigma'_j}^z,J_{ij} \to J \label{eq:2}.
\end{equation}
The Hamiltonian remains invariant after gauge transformation,
\begin{equation}
H =- \sum\limits_{ij} J_{ij}\sigma _i\sigma _j=- \sum\limits_{ij} J{\sigma'_i}^z{\sigma'_j}^z \label{eq:3}.
\end{equation}

\begin{figure}
\centerline{\includegraphics[width=3.0in]{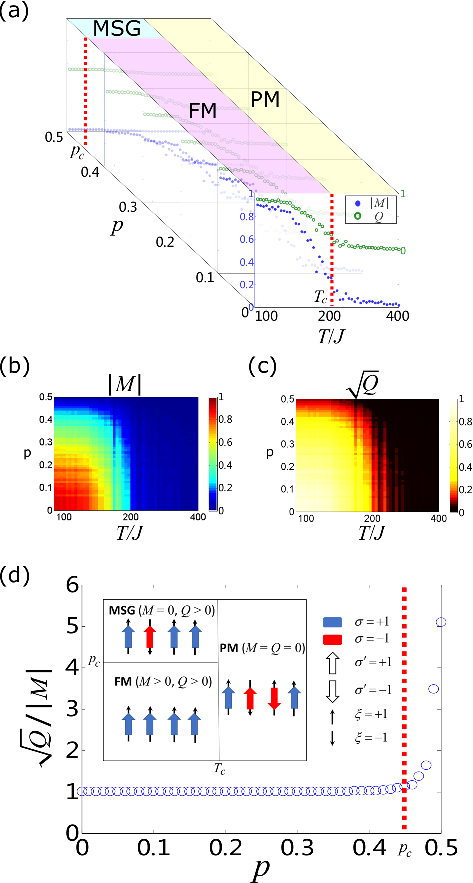}}
\caption{\label{fig:2}Phase diagram of the spatial spin glass system with 100 spins. (a) The absolute value of magnetization strength $M$ and spin glass order parameter $Q$ as functions of the effective temperature $T$ for fixed probabilities of $p=0,\text{ }0.1,\text{ }0.2,\text{ }0.3,\text{ }0.4,\text{ }0.5$. (b) The absolute value of magnetization strength $M$ and (c) the square-root value of spin glass order parameter $Q$ as functions of $p$ and $T$. The phase transition points are labeled by the critical temperature ${{T}_{c}}$ and the critical probability ${{p}_{c}}$. (d) The value of ${\sqrt{Q}}/{\left| M \right|}\;$ for spin glasses in low effective temperature $T=89.9J$, with respect to the probability $p$. The inset is the schematic of the spin configurations of three different phases: paramagnetic phase (PM), ferromagnetic phase (FM), and Mattis spin glass phase (MSG).}
\end{figure}

We experimentally implement the gauge transformation with the setup shown in Fig.~\ref{fig:1}(d). Instead of separately encoding the interaction strengths $J_{ij}=J\xi_i\xi_j$ and the original spin configuration $\mathbf{S}=\{\sigma_j\}$ on two SLMs, here we encode the gauge-transformed spin configuration ${\mathbf{S'}}^z=\{{\sigma'_j}^z\}$ on a single phase-only SLM. In this case, the rotation of spins corresponds to the modification of the phase modulation on SLM as
\begin{equation}
\varphi_{j,{\rm{SLM}}}=\sigma _j\frac{\pi }{2}+(-1)^j\alpha _j \label{eq:4}.
\end{equation}
By performing the gauge transformation, the requirement of amplitude modulation as shown in Fig.~\ref{fig:1}(b) is eliminated. The derivation of Eq. (\ref{eq:4}) is inspired by the complex encoding method with double-phase hologram \cite{hsueh1978computer,mendoza2014encoding,ngcobo2013digital,dudley2012controlling} and the details are given by SM Sec.~II. Due to the gauge invariance, the detected Hamiltonian after the gauge transformation remains invariant, as presented in Eq. (\ref{eq:3}).

\textit{Phase diagram for spatial photonic Ising machine.}---To show statistical equilibrium properties of spatial photonic Ising machine, we consider the interaction strengths ${{J}_{ij}}= \pm J$ \cite{nishimori2001statistical}, where each ${{\xi }_{j}}$ is randomly chosen following the distribution probability of
\begin{equation}
p(\xi_j)=p\cdot \delta (\xi_j,+1)+(1-p)\cdot \delta (\xi_j,-1) \label{eq:5}
\end{equation}
where $\delta (n,m)$ is a Kronecker delta function, i.e. each ${{\xi }_{j}}$ independently takes the value of either $+1$ or $-1$, with the probability of $p$ and $1-p$ respectively. With respect to the probability $p$ and an effective temperature $T$, the phases of this spin system are characterized by two statistical order parameters, the magnetization strength $M$ and the spin glass order parameter $Q$. These two statistical order parameters are defined as $M=\left[ \left\langle \frac{1}{N}\sum\limits_{j}{{{\sigma }_{j}}} \right\rangle  \right]$ and $Q=\left[ {{\left\langle \frac{1}{N}\sum\limits_{j}{{{\sigma }_{j}}} \right\rangle }^{2}} \right]$ respectively, where $\left\langle \cdots  \right\rangle $ denotes the ensemble average over the spin configurations $\mathbf{S}$ and $\left[ \cdots  \right]$ denotes the configurational average over different sets of $\{{{\xi }_{j}}\}$ generated following the probability $p$. In order to show the phase diagram of spatial spin glass system, we first formulate the statistical ensembles containing sufficient samples of spin configuration $\mathbf{S}$.

The proposed gauge transformation is rather convenient for investigating the phase transitions of these optical spin models. In this way, only one set of experiment is needed to compute the full phase diagram of the spin glass system. This is because though the interactions of the spin systems are different determined by different $\{{{\xi }_{j}}\}$, they share the same model after gauge transformation. In the experiment, it is the gauge-transformed effective spin configurations $\mathbf{{S}'}=\left\{ {\sigma'_j}^{z} \right\}$ that are encoded to formulate the statistical ensembles containing sufficient samples. With the knowledge of any given $\{{{\xi }_{j}}\}$, the statistical ensemble for the original spin configurations $\mathbf{S}$ before gauge transformation can be obtained by simply performing ${{\sigma }_{j}}={{\sigma'_j}^{z}}/{{{\xi }_{j}}}\;$, where for each given probability $p$, 100 different sets of $\{{{\xi }_{j}}\}$ are generated following the probability equation [Eq. (\ref{eq:5})] for configurational averaging in determining the order parameters $M$ and $Q$ of the spin glass systems.

In the experiment, an arbitrary initial effective spin configuration $\mathbf{{S}'}=\left\{ {\sigma'_j}^{z} \right\}$ with 100 spins, where each ${\sigma'_j}^z$ takes value of either $+1$ or $-1$ randomly, is encoded on the phase-only SLM (Holoeye PLUTO-NIR-011, $1920\times 1080$ pixels, with pixel size of $8\mu \text{m}\times 8\mu \text{m}$) following $\varphi_{j,\text{SLM}}\text{=}{\sigma'_j}^z\frac{\pi }{2}$. Here an active area with the size of $3200\mu \text{m}\times 3200\mu \text{m}$ on SLM is divided in to an array of 10-by-10 macropixels, where each macropixel encode an effective spin.

We generate the ensemble of equilibrium states in the spatial spin glass system with each given effective temperature $T$ following the measurement and feedback scheme. Governed by the Markov Chain Monte Carlo (MCMC) algorithm, we tentatively flip one spin on the SLM during each iteration and measure the normalized field intensity $I(u,v)$ on the back focal plane of the lens L1 (focal length $f=100\text{mm}$) by CCD (Ophir SP620), which then gives the Hamiltonian as Eq. (\ref{eq:1}). Following the Metropolis-Hasting sampling procedure (see SM Sec.~IV for details), with the effective temperature $T$, we update the spin configuration with the knowledge of the optically computed feedback Hamiltonian $H$. Under each fixed effective temperature $T$, 1000 Monte Carlo iterations are performed to formulate a statistical ensemble containing 1000 effective spin configurations $\{\mathbf{S'}_{1}^{(T)},\mathbf{{S}'}_{2}^{(T)},\cdots ,\mathbf{{S}'}_{1000}^{(T)}\}$. These 1000 samples represent the equilibrium states of the spin glass system, where the spin configurations satisfy the Gibbs-Boltzmann distribution, $p(\mathbf{S'})\propto e^{-H/T}$. We gradually cool the effective temperature from $T=400J$ to $T=89.9J$ in the simulated annealing manner. In this way, the whole statistical equilibrium states in the effective temperature range are obtained. The experimentally measured order parameters are presented in Fig.~\ref{fig:2}, forming the phase diagram with respect to $p$ and $T$. Here five experimental runs are conducted independently and then averaged to reduce the error. In general, following the processes described above, we can determine the full phase diagram of the spatial spin glass system with the aid of the gauge transformation.

As shown in Fig.~\ref{fig:2}, these two order parameters $M$ and $Q$ categorize the spin glass systems into three distinct phases, the paramagnetic phase (PM, $M=Q=0$), the ferromagnetic phase (FM, $M>0$, $Q>0$) and the Mattis spin glass phase (MSG, $M=0$, $Q>0$). Figure \ref{fig:2}(a) depicts the phase transition as a function of $T$ for fixed probabilities $p$. When $p=0$ for example, the spin system has uniform interactions, and it experiences the order-disorder phase transition, at a critical temperature $T_c=199J$, between the PM and FM phases during annealing. At high effective temperatures $T>T_c$, the original spins $\sigma_j$s align randomly, thus the averaged magnetization strength $M$ vanishes in the paramagnetic phase. At low effective temperatures $T<T_c$, the spins $\sigma_j$s tend to align in parallel to minimize the total energy $H$, resulting in spontaneous magnetizations with nonzero $M$ in the ferromagnetic phase. This phase transition between PM and FM during annealing still holds for other spin glass systems where the degree of disorder in interactions is lower than the critical probability $p_c$.

When $p>p_c$ however, a phase of MSG occurs at low effective temperatures, where the original spins $\sigma_j$s seem to be randomly distributed, but they are in fact locked to $\xi_j$s, that is, all $\sigma_j$s tend to align parallel to $\xi_j$s. Thus the gauge-transformed spins ${\sigma'_j}^{z}$s all take uniform value of $+1$ or $-1$ simultaneously, as shown in the inset of Fig.~\ref{fig:2}(d). The disorder in $\sigma_j$s originates from the disorder in the interaction strengths, i.e. the disorder of $\xi_j$s, which is different from the case in the PM phase. When the interaction strengths given by $\{\xi_j\}$ are fixed, there are no randomness in the values of $\{\sigma_j\}$ and the ensemble averaged magnetization strength $\left\langle \frac{1}{N}\sum\limits_{j}{\sigma_j} \right\rangle $ is non-vanishing ($Q>0$), while averaging over the disorder in interaction strengths then cancels the magnetization strength ($M=0$).

In order to determine the critical probability $p_c$, we distinguish the FM and MSG phases by plotting the values of $\sqrt{Q}/|M|$ with respect to the probability $p$, at a low effective temperature $T=89.9J$. The data are extracted from the left-most lines of Fig.~\ref{fig:2}(b) and (c). The results are shown in Fig.~\ref{fig:2}(d), and we determine $p_c=0.45$ where $\sqrt{Q}/|M|$ remains approximately 1 for $p<p_c$ and diverges when $p>p_c$, showing the characteristics of FM and MSG phases respectively. We note that the critical points  $T_c$ and $p_c$ determined from the phase diagram, agree well with the predictions of the mean-field theory, $T_c=2(N-1)J$ and $p_c^2+(1-p_c)^2={1}/(1+e^{-4J/{T_{c}}})$ \cite{nishimori2001statistical}. These results show that the approximation of the mean-field model is valid in the spatial photonic Ising machine.


\begin{figure}
\centerline{\includegraphics[width=3.0in]{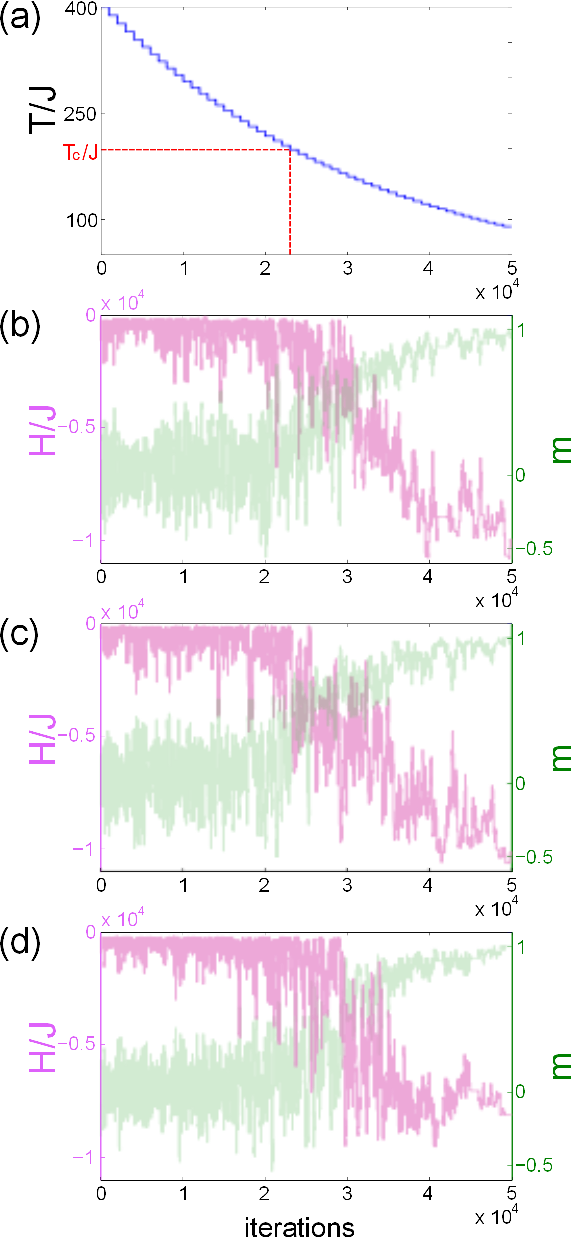}}
\caption{\label{fig:3}Experimental results for solving an optimization problem with $J_{ij}=+J$ for all the $N=100$ spins . (a) The system temperature is controlled with an exponentially decay during the Monte Carlo iterations, where $T_c$ corresponds to the critical temperature. (b-d) Three sets of measurements of the normalized Hamiltonian $H$ and the magnetization $ {{m}}$ during the cooling process.}
\end{figure}

\textit{Phase transistion in solving optimization problems.}--- The spatial spin glass system provides a new computation platform for solving the challenging combinatorial optimization problems by searching the ground state \cite{pierangeli2019large, pierangeli2020noise, pierangeli2020adiabatic, pierangeli2020scalable}. Since the process of searching ground states is in parallel with cooling spin systems, it is expected that the phase transitions should strongly impact the searching process. To show such impacts, here we consider solving a simple optimization problem where the spin interaction $J_{ij}=+J$ is a positive constant for all one-hundred spins. Figure~\ref{fig:3}(a) shows the temperature of the interacting system during the Monte Carlo iterations, which is controlled with an exponentially decay. The spins initially start from a random configuration, encoded on the SLM with an array of 10-by-10 macropixels.

Here the phases transition is also observed during the dynamical cooling process. Figures~\ref{fig:3}(b-d) show three sets of measurements of the normalized Hamiltonian $H$ and the magnetization $ {{m}= \frac{1}{N}\sum\limits_{i}{{\sigma_i}} }$ during the cooling process. All three experimental results demonstrate that the magnetization of the spins wanders around zero at high temperature, while the systems evolve to the ground state with $m=1$ by cooling. Such magnetizations indicate the transition from the paramagnetic phase to the ferromagnetic phase, and more importantly the phase-transition temperature coincides  with our previous determined $T_c$. More importantly, Figs.~\ref{fig:3}(b-d) also show strong fluctuations of $H$ in the vicinity of $T_c$, where the spin configuration coexists at two phases. Practically, it suggests that the spatial photonic Ising machine needs more iterations to return to equilibrium after perturbations in the vicinity of $T_c$. Below $T_c$, the system can be dramatically cooled down to the ground state. Since the system is associated with the optical aberrations and the intensity measurement uncertainty, the results show that the spatial photonic Ising machine is robust to such disturbance with sufficient many-spin interactions.

\textit{Discussion and Conclusion.}---Benefiting from the gauge transformation, we characterize the phase transition to understand the statistical spin dynamics in the novel optical spin system. We also notice a study to simulate large-scale random spin networks with a disordered medium \cite{pierangeli2020scalable}. However, the technique is distinct from the gauge transformation and cannot characterize the phase transition presented here. Furthermore, based on the Ising systems constructing with nonlinear processes as shown in \cite{kumar2020large}, we also believe that the proposed gauge transformation can be extended to study more complex spin models with four-body interactions.

In summary, we have proposed a spatial spin glass system with gauge transformation. The encoding of the spin configurations and the programming of the interaction strengths between spins are realized by only one spatial phase modulation process, which significantly improves the stability and fidelity of the optical Ising machine. For both studies of the statistical equilibrium state properties of spin glass systems and practical applications in solving combinatorial optimization problems, such an optical system exhibits high accuracy and great robustness against noises and aberrations, intriguing for its programmability and scalability in large-scale Ising machine.

The authors acknowledge funding through the National Natural Science Foundation of China (NSFC Grants Nos. 91850108 and 61675179), the National Key Research and Development
Program of China (Grant No. 2017YFA0205700), the Open Foundation of the State Key Laboratory of Modern Optical Instrumentation, and the Open Research Program of Key Laboratory of 3D Micro/Nano Fabrication and Characterization of Zhejiang Province.




\begin{thebibliography}{50}%
\makeatletter
\providecommand \@ifxundefined [1]{%
 \@ifx{#1\undefined}
}%
\providecommand \@ifnum [1]{%
 \ifnum #1\expandafter \@firstoftwo
 \else \expandafter \@secondoftwo
 \fi
}%
\providecommand \@ifx [1]{%
 \ifx #1\expandafter \@firstoftwo
 \else \expandafter \@secondoftwo
 \fi
}%
\providecommand \natexlab [1]{#1}%
\providecommand \enquote  [1]{``#1''}%
\providecommand \bibnamefont  [1]{#1}%
\providecommand \bibfnamefont [1]{#1}%
\providecommand \citenamefont [1]{#1}%
\providecommand \href@noop [0]{\@secondoftwo}%
\providecommand \href [0]{\begingroup \@sanitize@url \@href}%
\providecommand \@href[1]{\@@startlink{#1}\@@href}%
\providecommand \@@href[1]{\endgroup#1\@@endlink}%
\providecommand \@sanitize@url [0]{\catcode `\\12\catcode `\$12\catcode
  `\&12\catcode `\#12\catcode `\^12\catcode `\_12\catcode `\%12\relax}%
\providecommand \@@startlink[1]{}%
\providecommand \@@endlink[0]{}%
\providecommand \url  [0]{\begingroup\@sanitize@url \@url }%
\providecommand \@url [1]{\endgroup\@href {#1}{\urlprefix }}%
\providecommand \urlprefix  [0]{URL }%
\providecommand \Eprint [0]{\href }%
\providecommand \doibase [0]{https://doi.org/}%
\providecommand \selectlanguage [0]{\@gobble}%
\providecommand \bibinfo  [0]{\@secondoftwo}%
\providecommand \bibfield  [0]{\@secondoftwo}%
\providecommand \translation [1]{[#1]}%
\providecommand \BibitemOpen [0]{}%
\providecommand \bibitemStop [0]{}%
\providecommand \bibitemNoStop [0]{.\EOS\space}%
\providecommand \EOS [0]{\spacefactor3000\relax}%
\providecommand \BibitemShut  [1]{\csname bibitem#1\endcsname}%
\let\auto@bib@innerbib\@empty
\bibitem [{\citenamefont {Lucas}(2014)}]{lucas2014ising}%
  \BibitemOpen
  \bibfield  {author} {\bibinfo {author} {\bibfnamefont {A.}~\bibnamefont
  {Lucas}},\ }\bibfield  {title} {\bibinfo {title} {{Ising} formulations of
  many {NP} problems},\ }\href@noop {} {\bibfield  {journal} {\bibinfo
  {journal} {Frontiers in Physics}\ }\textbf {\bibinfo {volume} {2}},\ \bibinfo
  {pages} {5} (\bibinfo {year} {2014})}\BibitemShut {NoStop}%
\bibitem [{\citenamefont {McMahon}\ \emph {et~al.}(2016)\citenamefont
  {McMahon}, \citenamefont {Marandi}, \citenamefont {Haribara}, \citenamefont
  {Hamerly}, \citenamefont {Langrock}, \citenamefont {Tamate}, \citenamefont
  {Inagaki}, \citenamefont {Takesue}, \citenamefont {Utsunomiya}, \citenamefont
  {Aihara} \emph {et~al.}}]{mcmahon2016fully}%
  \BibitemOpen
  \bibfield  {author} {\bibinfo {author} {\bibfnamefont {P.~L.}\ \bibnamefont
  {McMahon}}, \bibinfo {author} {\bibfnamefont {A.}~\bibnamefont {Marandi}},
  \bibinfo {author} {\bibfnamefont {Y.}~\bibnamefont {Haribara}}, \bibinfo
  {author} {\bibfnamefont {R.}~\bibnamefont {Hamerly}}, \bibinfo {author}
  {\bibfnamefont {C.}~\bibnamefont {Langrock}}, \bibinfo {author}
  {\bibfnamefont {S.}~\bibnamefont {Tamate}}, \bibinfo {author} {\bibfnamefont
  {T.}~\bibnamefont {Inagaki}}, \bibinfo {author} {\bibfnamefont
  {H.}~\bibnamefont {Takesue}}, \bibinfo {author} {\bibfnamefont
  {S.}~\bibnamefont {Utsunomiya}}, \bibinfo {author} {\bibfnamefont
  {K.}~\bibnamefont {Aihara}}, \emph {et~al.},\ }\bibfield  {title} {\bibinfo
  {title} {A fully programmable 100-spin coherent {Ising} machine with
  all-to-all connections},\ }\href@noop {} {\bibfield  {journal} {\bibinfo
  {journal} {Science}\ }\textbf {\bibinfo {volume} {354}},\ \bibinfo {pages}
  {614} (\bibinfo {year} {2016})}\BibitemShut {NoStop}%
\bibitem [{\citenamefont {Inagaki}\ \emph
  {et~al.}(2016{\natexlab{a}})\citenamefont {Inagaki}, \citenamefont
  {Haribara}, \citenamefont {Igarashi}, \citenamefont {Sonobe}, \citenamefont
  {Tamate}, \citenamefont {Honjo}, \citenamefont {Marandi}, \citenamefont
  {McMahon}, \citenamefont {Umeki}, \citenamefont {Enbutsu} \emph
  {et~al.}}]{inagaki2016coherent}%
  \BibitemOpen
  \bibfield  {author} {\bibinfo {author} {\bibfnamefont {T.}~\bibnamefont
  {Inagaki}}, \bibinfo {author} {\bibfnamefont {Y.}~\bibnamefont {Haribara}},
  \bibinfo {author} {\bibfnamefont {K.}~\bibnamefont {Igarashi}}, \bibinfo
  {author} {\bibfnamefont {T.}~\bibnamefont {Sonobe}}, \bibinfo {author}
  {\bibfnamefont {S.}~\bibnamefont {Tamate}}, \bibinfo {author} {\bibfnamefont
  {T.}~\bibnamefont {Honjo}}, \bibinfo {author} {\bibfnamefont
  {A.}~\bibnamefont {Marandi}}, \bibinfo {author} {\bibfnamefont {P.~L.}\
  \bibnamefont {McMahon}}, \bibinfo {author} {\bibfnamefont {T.}~\bibnamefont
  {Umeki}}, \bibinfo {author} {\bibfnamefont {K.}~\bibnamefont {Enbutsu}},
  \emph {et~al.},\ }\bibfield  {title} {\bibinfo {title} {A coherent {Ising}
  machine for 2000-node optimization problems},\ }\href@noop {} {\bibfield
  {journal} {\bibinfo  {journal} {Science}\ }\textbf {\bibinfo {volume}
  {354}},\ \bibinfo {pages} {603} (\bibinfo {year}
  {2016}{\natexlab{a}})}\BibitemShut {NoStop}%
\bibitem [{\citenamefont {Inagaki}\ \emph
  {et~al.}(2016{\natexlab{b}})\citenamefont {Inagaki}, \citenamefont {Inaba},
  \citenamefont {Hamerly}, \citenamefont {Inoue}, \citenamefont {Yamamoto},\
  and\ \citenamefont {Takesue}}]{inagaki2016large}%
  \BibitemOpen
  \bibfield  {author} {\bibinfo {author} {\bibfnamefont {T.}~\bibnamefont
  {Inagaki}}, \bibinfo {author} {\bibfnamefont {K.}~\bibnamefont {Inaba}},
  \bibinfo {author} {\bibfnamefont {R.}~\bibnamefont {Hamerly}}, \bibinfo
  {author} {\bibfnamefont {K.}~\bibnamefont {Inoue}}, \bibinfo {author}
  {\bibfnamefont {Y.}~\bibnamefont {Yamamoto}},\ and\ \bibinfo {author}
  {\bibfnamefont {H.}~\bibnamefont {Takesue}},\ }\bibfield  {title} {\bibinfo
  {title} {Large-scale {Ising} spin network based on degenerate optical
  parametric oscillators},\ }\href@noop {} {\bibfield  {journal} {\bibinfo
  {journal} {Nature Photonics}\ }\textbf {\bibinfo {volume} {10}},\ \bibinfo
  {pages} {415} (\bibinfo {year} {2016}{\natexlab{b}})}\BibitemShut {NoStop}%
\bibitem [{\citenamefont {B{\"o}hm}\ \emph {et~al.}(2019)\citenamefont
  {B{\"o}hm}, \citenamefont {Verschaffelt},\ and\ \citenamefont {Van~der
  Sande}}]{bohm2019poor}%
  \BibitemOpen
  \bibfield  {author} {\bibinfo {author} {\bibfnamefont {F.}~\bibnamefont
  {B{\"o}hm}}, \bibinfo {author} {\bibfnamefont {G.}~\bibnamefont
  {Verschaffelt}},\ and\ \bibinfo {author} {\bibfnamefont {G.}~\bibnamefont
  {Van~der Sande}},\ }\bibfield  {title} {\bibinfo {title} {A poor man's
  coherent {Ising} machine based on opto-electronic feedback systems for
  solving optimization problems},\ }\href@noop {} {\bibfield  {journal}
  {\bibinfo  {journal} {Nature Communications}\ }\textbf {\bibinfo {volume}
  {10}},\ \bibinfo {pages} {1} (\bibinfo {year} {2019})}\BibitemShut {NoStop}%
\bibitem [{\citenamefont {Marandi}\ \emph {et~al.}(2014)\citenamefont
  {Marandi}, \citenamefont {Wang}, \citenamefont {Takata}, \citenamefont
  {Byer},\ and\ \citenamefont {Yamamoto}}]{marandi2014network}%
  \BibitemOpen
  \bibfield  {author} {\bibinfo {author} {\bibfnamefont {A.}~\bibnamefont
  {Marandi}}, \bibinfo {author} {\bibfnamefont {Z.}~\bibnamefont {Wang}},
  \bibinfo {author} {\bibfnamefont {K.}~\bibnamefont {Takata}}, \bibinfo
  {author} {\bibfnamefont {R.~L.}\ \bibnamefont {Byer}},\ and\ \bibinfo
  {author} {\bibfnamefont {Y.}~\bibnamefont {Yamamoto}},\ }\bibfield  {title}
  {\bibinfo {title} {Network of time-multiplexed optical parametric oscillators
  as a coherent {Ising} machine},\ }\href@noop {} {\bibfield  {journal}
  {\bibinfo  {journal} {Nature Photonics}\ }\textbf {\bibinfo {volume} {8}},\
  \bibinfo {pages} {937} (\bibinfo {year} {2014})}\BibitemShut {NoStop}%
\bibitem [{\citenamefont {Utsunomiya}\ \emph {et~al.}(2011)\citenamefont
  {Utsunomiya}, \citenamefont {Takata},\ and\ \citenamefont
  {Yamamoto}}]{utsunomiya2011mapping}%
  \BibitemOpen
  \bibfield  {author} {\bibinfo {author} {\bibfnamefont {S.}~\bibnamefont
  {Utsunomiya}}, \bibinfo {author} {\bibfnamefont {K.}~\bibnamefont {Takata}},\
  and\ \bibinfo {author} {\bibfnamefont {Y.}~\bibnamefont {Yamamoto}},\
  }\bibfield  {title} {\bibinfo {title} {Mapping of {Ising} models onto
  injection-locked laser systems},\ }\href@noop {} {\bibfield  {journal}
  {\bibinfo  {journal} {Optics Express}\ }\textbf {\bibinfo {volume} {19}},\
  \bibinfo {pages} {18091} (\bibinfo {year} {2011})}\BibitemShut {NoStop}%
\bibitem [{\citenamefont {Babaeian}\ \emph {et~al.}(2019)\citenamefont
  {Babaeian}, \citenamefont {Nguyen}, \citenamefont {Demir}, \citenamefont
  {Akbulut}, \citenamefont {Blanche}, \citenamefont {Kaneda}, \citenamefont
  {Guha}, \citenamefont {Neifeld},\ and\ \citenamefont
  {Peyghambarian}}]{babaeian2019single}%
  \BibitemOpen
  \bibfield  {author} {\bibinfo {author} {\bibfnamefont {M.}~\bibnamefont
  {Babaeian}}, \bibinfo {author} {\bibfnamefont {D.~T.}\ \bibnamefont
  {Nguyen}}, \bibinfo {author} {\bibfnamefont {V.}~\bibnamefont {Demir}},
  \bibinfo {author} {\bibfnamefont {M.}~\bibnamefont {Akbulut}}, \bibinfo
  {author} {\bibfnamefont {P.-A.}\ \bibnamefont {Blanche}}, \bibinfo {author}
  {\bibfnamefont {Y.}~\bibnamefont {Kaneda}}, \bibinfo {author} {\bibfnamefont
  {S.}~\bibnamefont {Guha}}, \bibinfo {author} {\bibfnamefont {M.~A.}\
  \bibnamefont {Neifeld}},\ and\ \bibinfo {author} {\bibfnamefont
  {N.}~\bibnamefont {Peyghambarian}},\ }\bibfield  {title} {\bibinfo {title} {A
  single shot coherent {Ising} machine based on a network of injection-locked
  multicore fiber lasers},\ }\href@noop {} {\bibfield  {journal} {\bibinfo
  {journal} {Nature Communications}\ }\textbf {\bibinfo {volume} {10}},\
  \bibinfo {pages} {1} (\bibinfo {year} {2019})}\BibitemShut {NoStop}%
\bibitem [{\citenamefont {Tradonsky}\ \emph {et~al.}(2019)\citenamefont
  {Tradonsky}, \citenamefont {Gershenzon}, \citenamefont {Pal}, \citenamefont
  {Chriki}, \citenamefont {Friesem}, \citenamefont {Raz},\ and\ \citenamefont
  {Davidson}}]{tradonsky2019rapid}%
  \BibitemOpen
  \bibfield  {author} {\bibinfo {author} {\bibfnamefont {C.}~\bibnamefont
  {Tradonsky}}, \bibinfo {author} {\bibfnamefont {I.}~\bibnamefont
  {Gershenzon}}, \bibinfo {author} {\bibfnamefont {V.}~\bibnamefont {Pal}},
  \bibinfo {author} {\bibfnamefont {R.}~\bibnamefont {Chriki}}, \bibinfo
  {author} {\bibfnamefont {A.~A.}\ \bibnamefont {Friesem}}, \bibinfo {author}
  {\bibfnamefont {O.}~\bibnamefont {Raz}},\ and\ \bibinfo {author}
  {\bibfnamefont {N.}~\bibnamefont {Davidson}},\ }\bibfield  {title} {\bibinfo
  {title} {Rapid laser solver for the phase retrieval problem},\ }\href@noop {}
  {\bibfield  {journal} {\bibinfo  {journal} {Science Advances}\ }\textbf
  {\bibinfo {volume} {5}},\ \bibinfo {pages} {eaax4530} (\bibinfo {year}
  {2019})}\BibitemShut {NoStop}%
\bibitem [{\citenamefont {Parto}\ \emph {et~al.}(2020)\citenamefont {Parto},
  \citenamefont {Hayenga}, \citenamefont {Marandi}, \citenamefont
  {Christodoulides},\ and\ \citenamefont {Khajavikhan}}]{parto2020realizing}%
  \BibitemOpen
  \bibfield  {author} {\bibinfo {author} {\bibfnamefont {M.}~\bibnamefont
  {Parto}}, \bibinfo {author} {\bibfnamefont {W.}~\bibnamefont {Hayenga}},
  \bibinfo {author} {\bibfnamefont {A.}~\bibnamefont {Marandi}}, \bibinfo
  {author} {\bibfnamefont {D.~N.}\ \bibnamefont {Christodoulides}},\ and\
  \bibinfo {author} {\bibfnamefont {M.}~\bibnamefont {Khajavikhan}},\
  }\bibfield  {title} {\bibinfo {title} {Realizing spin {Hamiltonians} in
  nanoscale active photonic lattices},\ }\href@noop {} {\bibfield  {journal}
  {\bibinfo  {journal} {Nature Materials}\ }\textbf {\bibinfo {volume} {19}},\
  \bibinfo {pages} {725} (\bibinfo {year} {2020})}\BibitemShut {NoStop}%
\bibitem [{\citenamefont {Honari-Latifpour}\ and\ \citenamefont
  {Miri}(2020)}]{honari2020mapping}%
  \BibitemOpen
  \bibfield  {author} {\bibinfo {author} {\bibfnamefont {M.}~\bibnamefont
  {Honari-Latifpour}}\ and\ \bibinfo {author} {\bibfnamefont {M.-A.}\
  \bibnamefont {Miri}},\ }\bibfield  {title} {\bibinfo {title} {Mapping the
  {XY} {Hamiltonian} onto a network of coupled lasers},\ }\href@noop {}
  {\bibfield  {journal} {\bibinfo  {journal} {Physical Review Research}\
  }\textbf {\bibinfo {volume} {2}},\ \bibinfo {pages} {043335} (\bibinfo {year}
  {2020})}\BibitemShut {NoStop}%
\bibitem [{\citenamefont {Kalinin}\ \emph {et~al.}(2020)\citenamefont
  {Kalinin}, \citenamefont {Amo}, \citenamefont {Bloch},\ and\ \citenamefont
  {Berloff}}]{kalinin2020polaritonic}%
  \BibitemOpen
  \bibfield  {author} {\bibinfo {author} {\bibfnamefont {K.~P.}\ \bibnamefont
  {Kalinin}}, \bibinfo {author} {\bibfnamefont {A.}~\bibnamefont {Amo}},
  \bibinfo {author} {\bibfnamefont {J.}~\bibnamefont {Bloch}},\ and\ \bibinfo
  {author} {\bibfnamefont {N.~G.}\ \bibnamefont {Berloff}},\ }\bibfield
  {title} {\bibinfo {title} {Polaritonic {XY-Ising} machine},\ }\href@noop {}
  {\bibfield  {journal} {\bibinfo  {journal} {Nanophotonics}\ }\textbf
  {\bibinfo {volume} {9}},\ \bibinfo {pages} {4127} (\bibinfo {year}
  {2020})}\BibitemShut {NoStop}%
\bibitem [{\citenamefont {Berloff}\ \emph {et~al.}(2017)\citenamefont
  {Berloff}, \citenamefont {Silva}, \citenamefont {Kalinin}, \citenamefont
  {Askitopoulos}, \citenamefont {T{\"o}pfer}, \citenamefont {Cilibrizzi},
  \citenamefont {Langbein},\ and\ \citenamefont
  {Lagoudakis}}]{berloff2017realizing}%
  \BibitemOpen
  \bibfield  {author} {\bibinfo {author} {\bibfnamefont {N.~G.}\ \bibnamefont
  {Berloff}}, \bibinfo {author} {\bibfnamefont {M.}~\bibnamefont {Silva}},
  \bibinfo {author} {\bibfnamefont {K.}~\bibnamefont {Kalinin}}, \bibinfo
  {author} {\bibfnamefont {A.}~\bibnamefont {Askitopoulos}}, \bibinfo {author}
  {\bibfnamefont {J.~D.}\ \bibnamefont {T{\"o}pfer}}, \bibinfo {author}
  {\bibfnamefont {P.}~\bibnamefont {Cilibrizzi}}, \bibinfo {author}
  {\bibfnamefont {W.}~\bibnamefont {Langbein}},\ and\ \bibinfo {author}
  {\bibfnamefont {P.~G.}\ \bibnamefont {Lagoudakis}},\ }\bibfield  {title}
  {\bibinfo {title} {Realizing the classical {XY} {Hamiltonian} in polariton
  simulators},\ }\href@noop {} {\bibfield  {journal} {\bibinfo  {journal}
  {Nature Materials}\ }\textbf {\bibinfo {volume} {16}},\ \bibinfo {pages}
  {1120} (\bibinfo {year} {2017})}\BibitemShut {NoStop}%
\bibitem [{\citenamefont {Kalinin}\ and\ \citenamefont
  {Berloff}(2018)}]{kalinin2018simulating}%
  \BibitemOpen
  \bibfield  {author} {\bibinfo {author} {\bibfnamefont {K.~P.}\ \bibnamefont
  {Kalinin}}\ and\ \bibinfo {author} {\bibfnamefont {N.~G.}\ \bibnamefont
  {Berloff}},\ }\bibfield  {title} {\bibinfo {title} {Simulating {Ising} and
  n-state planar potts models and external fields with nonequilibrium
  condensates},\ }\href@noop {} {\bibfield  {journal} {\bibinfo  {journal}
  {Physical Review Letters}\ }\textbf {\bibinfo {volume} {121}},\ \bibinfo
  {pages} {235302} (\bibinfo {year} {2018})}\BibitemShut {NoStop}%
\bibitem [{\citenamefont {Kim}\ \emph {et~al.}(2010)\citenamefont {Kim},
  \citenamefont {Chang}, \citenamefont {Korenblit}, \citenamefont {Islam},
  \citenamefont {Edwards}, \citenamefont {Freericks}, \citenamefont {Lin},
  \citenamefont {Duan},\ and\ \citenamefont {Monroe}}]{kim2010quantum}%
  \BibitemOpen
  \bibfield  {author} {\bibinfo {author} {\bibfnamefont {K.}~\bibnamefont
  {Kim}}, \bibinfo {author} {\bibfnamefont {M.-S.}\ \bibnamefont {Chang}},
  \bibinfo {author} {\bibfnamefont {S.}~\bibnamefont {Korenblit}}, \bibinfo
  {author} {\bibfnamefont {R.}~\bibnamefont {Islam}}, \bibinfo {author}
  {\bibfnamefont {E.~E.}\ \bibnamefont {Edwards}}, \bibinfo {author}
  {\bibfnamefont {J.~K.}\ \bibnamefont {Freericks}}, \bibinfo {author}
  {\bibfnamefont {G.-D.}\ \bibnamefont {Lin}}, \bibinfo {author} {\bibfnamefont
  {L.-M.}\ \bibnamefont {Duan}},\ and\ \bibinfo {author} {\bibfnamefont
  {C.}~\bibnamefont {Monroe}},\ }\bibfield  {title} {\bibinfo {title} {Quantum
  simulation of frustrated {Ising} spins with trapped ions},\ }\href@noop {}
  {\bibfield  {journal} {\bibinfo  {journal} {Nature}\ }\textbf {\bibinfo
  {volume} {465}},\ \bibinfo {pages} {590} (\bibinfo {year}
  {2010})}\BibitemShut {NoStop}%
\bibitem [{\citenamefont {Struck}\ \emph {et~al.}(2013)\citenamefont {Struck},
  \citenamefont {Weinberg}, \citenamefont {{\"O}lschl{\"a}ger}, \citenamefont
  {Windpassinger}, \citenamefont {Simonet}, \citenamefont {Sengstock},
  \citenamefont {H{\"o}ppner}, \citenamefont {Hauke}, \citenamefont {Eckardt},
  \citenamefont {Lewenstein} \emph {et~al.}}]{struck2013engineering}%
  \BibitemOpen
  \bibfield  {author} {\bibinfo {author} {\bibfnamefont {J.}~\bibnamefont
  {Struck}}, \bibinfo {author} {\bibfnamefont {M.}~\bibnamefont {Weinberg}},
  \bibinfo {author} {\bibfnamefont {C.}~\bibnamefont {{\"O}lschl{\"a}ger}},
  \bibinfo {author} {\bibfnamefont {P.}~\bibnamefont {Windpassinger}}, \bibinfo
  {author} {\bibfnamefont {J.}~\bibnamefont {Simonet}}, \bibinfo {author}
  {\bibfnamefont {K.}~\bibnamefont {Sengstock}}, \bibinfo {author}
  {\bibfnamefont {R.}~\bibnamefont {H{\"o}ppner}}, \bibinfo {author}
  {\bibfnamefont {P.}~\bibnamefont {Hauke}}, \bibinfo {author} {\bibfnamefont
  {A.}~\bibnamefont {Eckardt}}, \bibinfo {author} {\bibfnamefont
  {M.}~\bibnamefont {Lewenstein}}, \emph {et~al.},\ }\bibfield  {title}
  {\bibinfo {title} {Engineering {Ising-XY} spin-models in a triangular lattice
  using tunable artificial gauge fields},\ }\href@noop {} {\bibfield  {journal}
  {\bibinfo  {journal} {Nature Physics}\ }\textbf {\bibinfo {volume} {9}},\
  \bibinfo {pages} {738} (\bibinfo {year} {2013})}\BibitemShut {NoStop}%
\bibitem [{\citenamefont {Kassenberg}\ \emph {et~al.}(2020)\citenamefont
  {Kassenberg}, \citenamefont {Vretenar}, \citenamefont {Bissesar},\ and\
  \citenamefont {Klaers}}]{kassenberg2020controllable}%
  \BibitemOpen
  \bibfield  {author} {\bibinfo {author} {\bibfnamefont {B.}~\bibnamefont
  {Kassenberg}}, \bibinfo {author} {\bibfnamefont {M.}~\bibnamefont
  {Vretenar}}, \bibinfo {author} {\bibfnamefont {S.}~\bibnamefont {Bissesar}},\
  and\ \bibinfo {author} {\bibfnamefont {J.}~\bibnamefont {Klaers}},\
  }\bibfield  {title} {\bibinfo {title} {Controllable {Josephson} junction for
  photon {Bose--Einstein} condensates},\ }\href@noop {} {\bibfield  {journal}
  {\bibinfo  {journal} {arXiv preprint arXiv:2001.09828}\ } (\bibinfo {year}
  {2020})}\BibitemShut {NoStop}%
\bibitem [{\citenamefont {Cai}\ \emph {et~al.}(2020)\citenamefont {Cai},
  \citenamefont {Kumar}, \citenamefont {Van~Vaerenbergh}, \citenamefont
  {Sheng}, \citenamefont {Liu}, \citenamefont {Li}, \citenamefont {Liu},
  \citenamefont {Foltin}, \citenamefont {Yu}, \citenamefont {Xia},
  \citenamefont {Yang}, \citenamefont {Beausoleil}, \citenamefont {Lu},\ and\
  \citenamefont {Strachan}}]{cai2020power}%
  \BibitemOpen
  \bibfield  {author} {\bibinfo {author} {\bibfnamefont {F.}~\bibnamefont
  {Cai}}, \bibinfo {author} {\bibfnamefont {S.}~\bibnamefont {Kumar}}, \bibinfo
  {author} {\bibfnamefont {T.}~\bibnamefont {Van~Vaerenbergh}}, \bibinfo
  {author} {\bibfnamefont {X.}~\bibnamefont {Sheng}}, \bibinfo {author}
  {\bibfnamefont {R.}~\bibnamefont {Liu}}, \bibinfo {author} {\bibfnamefont
  {C.}~\bibnamefont {Li}}, \bibinfo {author} {\bibfnamefont {Z.}~\bibnamefont
  {Liu}}, \bibinfo {author} {\bibfnamefont {M.}~\bibnamefont {Foltin}},
  \bibinfo {author} {\bibfnamefont {S.}~\bibnamefont {Yu}}, \bibinfo {author}
  {\bibfnamefont {Q.}~\bibnamefont {Xia}}, \bibinfo {author} {\bibfnamefont
  {J.~J.}\ \bibnamefont {Yang}}, \bibinfo {author} {\bibfnamefont
  {R.}~\bibnamefont {Beausoleil}}, \bibinfo {author} {\bibfnamefont {W.~D.}\
  \bibnamefont {Lu}},\ and\ \bibinfo {author} {\bibfnamefont {J.~P.}\
  \bibnamefont {Strachan}},\ }\bibfield  {title} {\bibinfo {title}
  {Power-efficient combinatorial optimization using intrinsic noise in
  memristor hopfield neural networks},\ }\href@noop {} {\bibfield  {journal}
  {\bibinfo  {journal} {Nature Electronics}\ }\textbf {\bibinfo {volume} {3}},\
  \bibinfo {pages} {409} (\bibinfo {year} {2020})}\BibitemShut {NoStop}%
\bibitem [{\citenamefont {Johnson}\ \emph {et~al.}(2011)\citenamefont
  {Johnson}, \citenamefont {Amin}, \citenamefont {Gildert}, \citenamefont
  {Lanting}, \citenamefont {Hamze}, \citenamefont {Dickson}, \citenamefont
  {Harris}, \citenamefont {Berkley}, \citenamefont {Johansson}, \citenamefont
  {Bunyk} \emph {et~al.}}]{johnson2011quantum}%
  \BibitemOpen
  \bibfield  {author} {\bibinfo {author} {\bibfnamefont {M.~W.}\ \bibnamefont
  {Johnson}}, \bibinfo {author} {\bibfnamefont {M.~H.~S.}\ \bibnamefont
  {Amin}}, \bibinfo {author} {\bibfnamefont {S.}~\bibnamefont {Gildert}},
  \bibinfo {author} {\bibfnamefont {T.}~\bibnamefont {Lanting}}, \bibinfo
  {author} {\bibfnamefont {F.}~\bibnamefont {Hamze}}, \bibinfo {author}
  {\bibfnamefont {N.}~\bibnamefont {Dickson}}, \bibinfo {author} {\bibfnamefont
  {R.}~\bibnamefont {Harris}}, \bibinfo {author} {\bibfnamefont {A.~J.}\
  \bibnamefont {Berkley}}, \bibinfo {author} {\bibfnamefont {J.}~\bibnamefont
  {Johansson}}, \bibinfo {author} {\bibfnamefont {P.}~\bibnamefont {Bunyk}},
  \emph {et~al.},\ }\bibfield  {title} {\bibinfo {title} {Quantum annealing
  with manufactured spins},\ }\href@noop {} {\bibfield  {journal} {\bibinfo
  {journal} {Nature}\ }\textbf {\bibinfo {volume} {473}},\ \bibinfo {pages}
  {194} (\bibinfo {year} {2011})}\BibitemShut {NoStop}%
\bibitem [{\citenamefont {Boixo}\ \emph {et~al.}(2014)\citenamefont {Boixo},
  \citenamefont {R{\o}nnow}, \citenamefont {Isakov}, \citenamefont {Wang},
  \citenamefont {Wecker}, \citenamefont {Lidar}, \citenamefont {Martinis},\
  and\ \citenamefont {Troyer}}]{boixo2014evidence}%
  \BibitemOpen
  \bibfield  {author} {\bibinfo {author} {\bibfnamefont {S.}~\bibnamefont
  {Boixo}}, \bibinfo {author} {\bibfnamefont {T.~F.}\ \bibnamefont
  {R{\o}nnow}}, \bibinfo {author} {\bibfnamefont {S.~V.}\ \bibnamefont
  {Isakov}}, \bibinfo {author} {\bibfnamefont {Z.}~\bibnamefont {Wang}},
  \bibinfo {author} {\bibfnamefont {D.}~\bibnamefont {Wecker}}, \bibinfo
  {author} {\bibfnamefont {D.~A.}\ \bibnamefont {Lidar}}, \bibinfo {author}
  {\bibfnamefont {J.~M.}\ \bibnamefont {Martinis}},\ and\ \bibinfo {author}
  {\bibfnamefont {M.}~\bibnamefont {Troyer}},\ }\bibfield  {title} {\bibinfo
  {title} {Evidence for quantum annealing with more than one hundred qubits},\
  }\href@noop {} {\bibfield  {journal} {\bibinfo  {journal} {Nature Physics}\
  }\textbf {\bibinfo {volume} {10}},\ \bibinfo {pages} {218} (\bibinfo {year}
  {2014})}\BibitemShut {NoStop}%
\bibitem [{\citenamefont {King}\ \emph {et~al.}(2018)\citenamefont {King},
  \citenamefont {Carrasquilla}, \citenamefont {Raymond}, \citenamefont
  {Ozfidan}, \citenamefont {Andriyash}, \citenamefont {Berkley}, \citenamefont
  {Reis}, \citenamefont {Lanting}, \citenamefont {Harris}, \citenamefont
  {Altomare} \emph {et~al.}}]{king2018observation}%
  \BibitemOpen
  \bibfield  {author} {\bibinfo {author} {\bibfnamefont {A.~D.}\ \bibnamefont
  {King}}, \bibinfo {author} {\bibfnamefont {J.}~\bibnamefont {Carrasquilla}},
  \bibinfo {author} {\bibfnamefont {J.}~\bibnamefont {Raymond}}, \bibinfo
  {author} {\bibfnamefont {I.}~\bibnamefont {Ozfidan}}, \bibinfo {author}
  {\bibfnamefont {E.}~\bibnamefont {Andriyash}}, \bibinfo {author}
  {\bibfnamefont {A.}~\bibnamefont {Berkley}}, \bibinfo {author} {\bibfnamefont
  {M.}~\bibnamefont {Reis}}, \bibinfo {author} {\bibfnamefont {T.}~\bibnamefont
  {Lanting}}, \bibinfo {author} {\bibfnamefont {R.}~\bibnamefont {Harris}},
  \bibinfo {author} {\bibfnamefont {F.}~\bibnamefont {Altomare}}, \emph
  {et~al.},\ }\bibfield  {title} {\bibinfo {title} {Observation of topological
  phenomena in a programmable lattice of 1,800 qubits},\ }\href@noop {}
  {\bibfield  {journal} {\bibinfo  {journal} {Nature}\ }\textbf {\bibinfo
  {volume} {560}},\ \bibinfo {pages} {456} (\bibinfo {year}
  {2018})}\BibitemShut {NoStop}%
\bibitem [{\citenamefont {Roques-Carmes}\ \emph {et~al.}(2020)\citenamefont
  {Roques-Carmes}, \citenamefont {Shen}, \citenamefont {Zanoci}, \citenamefont
  {Prabhu}, \citenamefont {Atieh}, \citenamefont {Jing}, \citenamefont
  {Dub{\v{c}}ek}, \citenamefont {Mao}, \citenamefont {Johnson}, \citenamefont
  {{\v{C}}eperi{\'c}} \emph {et~al.}}]{roques2020heuristic}%
  \BibitemOpen
  \bibfield  {author} {\bibinfo {author} {\bibfnamefont {C.}~\bibnamefont
  {Roques-Carmes}}, \bibinfo {author} {\bibfnamefont {Y.}~\bibnamefont {Shen}},
  \bibinfo {author} {\bibfnamefont {C.}~\bibnamefont {Zanoci}}, \bibinfo
  {author} {\bibfnamefont {M.}~\bibnamefont {Prabhu}}, \bibinfo {author}
  {\bibfnamefont {F.}~\bibnamefont {Atieh}}, \bibinfo {author} {\bibfnamefont
  {L.}~\bibnamefont {Jing}}, \bibinfo {author} {\bibfnamefont {T.}~\bibnamefont
  {Dub{\v{c}}ek}}, \bibinfo {author} {\bibfnamefont {C.}~\bibnamefont {Mao}},
  \bibinfo {author} {\bibfnamefont {M.~R.}\ \bibnamefont {Johnson}}, \bibinfo
  {author} {\bibfnamefont {V.}~\bibnamefont {{\v{C}}eperi{\'c}}}, \emph
  {et~al.},\ }\bibfield  {title} {\bibinfo {title} {Heuristic recurrent
  algorithms for photonic {Ising} machines},\ }\href@noop {} {\bibfield
  {journal} {\bibinfo  {journal} {Nature Communications}\ }\textbf {\bibinfo
  {volume} {11}} (\bibinfo {year} {2020})}\BibitemShut {NoStop}%
\bibitem [{\citenamefont {Prabhu}\ \emph {et~al.}(2020)\citenamefont {Prabhu},
  \citenamefont {Roques-Carmes}, \citenamefont {Shen}, \citenamefont {Harris},
  \citenamefont {Jing}, \citenamefont {Carolan}, \citenamefont {Hamerly},
  \citenamefont {Baehr-Jones}, \citenamefont {Hochberg}, \citenamefont
  {{\v{C}}eperi{\'c}} \emph {et~al.}}]{prabhu2020accelerating}%
  \BibitemOpen
  \bibfield  {author} {\bibinfo {author} {\bibfnamefont {M.}~\bibnamefont
  {Prabhu}}, \bibinfo {author} {\bibfnamefont {C.}~\bibnamefont
  {Roques-Carmes}}, \bibinfo {author} {\bibfnamefont {Y.}~\bibnamefont {Shen}},
  \bibinfo {author} {\bibfnamefont {N.}~\bibnamefont {Harris}}, \bibinfo
  {author} {\bibfnamefont {L.}~\bibnamefont {Jing}}, \bibinfo {author}
  {\bibfnamefont {J.}~\bibnamefont {Carolan}}, \bibinfo {author} {\bibfnamefont
  {R.}~\bibnamefont {Hamerly}}, \bibinfo {author} {\bibfnamefont
  {T.}~\bibnamefont {Baehr-Jones}}, \bibinfo {author} {\bibfnamefont
  {M.}~\bibnamefont {Hochberg}}, \bibinfo {author} {\bibfnamefont
  {V.}~\bibnamefont {{\v{C}}eperi{\'c}}}, \emph {et~al.},\ }\bibfield  {title}
  {\bibinfo {title} {Accelerating recurrent {Ising} machines in photonic
  integrated circuits},\ }\href@noop {} {\bibfield  {journal} {\bibinfo
  {journal} {Optica}\ }\textbf {\bibinfo {volume} {7}},\ \bibinfo {pages} {551}
  (\bibinfo {year} {2020})}\BibitemShut {NoStop}%
\bibitem [{\citenamefont {Shen}\ \emph {et~al.}(2017)\citenamefont {Shen},
  \citenamefont {Harris}, \citenamefont {Skirlo}, \citenamefont {Prabhu},
  \citenamefont {Baehr-Jones}, \citenamefont {Hochberg}, \citenamefont {Sun},
  \citenamefont {Zhao}, \citenamefont {Larochelle}, \citenamefont {Englund}
  \emph {et~al.}}]{shen2017deep}%
  \BibitemOpen
  \bibfield  {author} {\bibinfo {author} {\bibfnamefont {Y.}~\bibnamefont
  {Shen}}, \bibinfo {author} {\bibfnamefont {N.~C.}\ \bibnamefont {Harris}},
  \bibinfo {author} {\bibfnamefont {S.}~\bibnamefont {Skirlo}}, \bibinfo
  {author} {\bibfnamefont {M.}~\bibnamefont {Prabhu}}, \bibinfo {author}
  {\bibfnamefont {T.}~\bibnamefont {Baehr-Jones}}, \bibinfo {author}
  {\bibfnamefont {M.}~\bibnamefont {Hochberg}}, \bibinfo {author}
  {\bibfnamefont {X.}~\bibnamefont {Sun}}, \bibinfo {author} {\bibfnamefont
  {S.}~\bibnamefont {Zhao}}, \bibinfo {author} {\bibfnamefont {H.}~\bibnamefont
  {Larochelle}}, \bibinfo {author} {\bibfnamefont {D.}~\bibnamefont {Englund}},
  \emph {et~al.},\ }\bibfield  {title} {\bibinfo {title} {Deep learning with
  coherent nanophotonic circuits},\ }\href@noop {} {\bibfield  {journal}
  {\bibinfo  {journal} {Nature Photonics}\ }\textbf {\bibinfo {volume} {11}},\
  \bibinfo {pages} {441} (\bibinfo {year} {2017})}\BibitemShut {NoStop}%
\bibitem [{\citenamefont {Wu}\ \emph {et~al.}(2014)\citenamefont {Wu},
  \citenamefont {De~Abajo}, \citenamefont {Soci}, \citenamefont {Shum},\ and\
  \citenamefont {Zheludev}}]{wu2014optical}%
  \BibitemOpen
  \bibfield  {author} {\bibinfo {author} {\bibfnamefont {K.}~\bibnamefont
  {Wu}}, \bibinfo {author} {\bibfnamefont {J.~G.}\ \bibnamefont {De~Abajo}},
  \bibinfo {author} {\bibfnamefont {C.}~\bibnamefont {Soci}}, \bibinfo {author}
  {\bibfnamefont {P.~P.}\ \bibnamefont {Shum}},\ and\ \bibinfo {author}
  {\bibfnamefont {N.~I.}\ \bibnamefont {Zheludev}},\ }\bibfield  {title}
  {\bibinfo {title} {An optical fiber network oracle for {NP}-complete
  problems},\ }\href@noop {} {\bibfield  {journal} {\bibinfo  {journal} {Light:
  Science \& Applications}\ }\textbf {\bibinfo {volume} {3}},\ \bibinfo {pages}
  {e147} (\bibinfo {year} {2014})}\BibitemShut {NoStop}%
\bibitem [{\citenamefont {Okawachi}\ \emph {et~al.}(2020)\citenamefont
  {Okawachi}, \citenamefont {Yu}, \citenamefont {Jang}, \citenamefont {Ji},
  \citenamefont {Zhao}, \citenamefont {Kim}, \citenamefont {Lipson},\ and\
  \citenamefont {Gaeta}}]{okawachi2020demonstration}%
  \BibitemOpen
  \bibfield  {author} {\bibinfo {author} {\bibfnamefont {Y.}~\bibnamefont
  {Okawachi}}, \bibinfo {author} {\bibfnamefont {M.}~\bibnamefont {Yu}},
  \bibinfo {author} {\bibfnamefont {J.~K.}\ \bibnamefont {Jang}}, \bibinfo
  {author} {\bibfnamefont {X.}~\bibnamefont {Ji}}, \bibinfo {author}
  {\bibfnamefont {Y.}~\bibnamefont {Zhao}}, \bibinfo {author} {\bibfnamefont
  {B.~Y.}\ \bibnamefont {Kim}}, \bibinfo {author} {\bibfnamefont
  {M.}~\bibnamefont {Lipson}},\ and\ \bibinfo {author} {\bibfnamefont {A.~L.}\
  \bibnamefont {Gaeta}},\ }\bibfield  {title} {\bibinfo {title} {Demonstration
  of chip-based coupled degenerate optical parametric oscillators for realizing
  a nanophotonic spin-glass},\ }\href@noop {} {\bibfield  {journal} {\bibinfo
  {journal} {Nature Communications}\ }\textbf {\bibinfo {volume} {11}},\
  \bibinfo {pages} {1} (\bibinfo {year} {2020})}\BibitemShut {NoStop}%
\bibitem [{\citenamefont {Pierangeli}\ \emph {et~al.}(2019)\citenamefont
  {Pierangeli}, \citenamefont {Marcucci},\ and\ \citenamefont
  {Conti}}]{pierangeli2019large}%
  \BibitemOpen
  \bibfield  {author} {\bibinfo {author} {\bibfnamefont {D.}~\bibnamefont
  {Pierangeli}}, \bibinfo {author} {\bibfnamefont {G.}~\bibnamefont
  {Marcucci}},\ and\ \bibinfo {author} {\bibfnamefont {C.}~\bibnamefont
  {Conti}},\ }\bibfield  {title} {\bibinfo {title} {Large-scale photonic
  {Ising} machine by spatial light modulation},\ }\href@noop {} {\bibfield
  {journal} {\bibinfo  {journal} {Physical Review Letters}\ }\textbf {\bibinfo
  {volume} {122}},\ \bibinfo {pages} {213902} (\bibinfo {year}
  {2019})}\BibitemShut {NoStop}%
\bibitem [{\citenamefont {Silva}\ \emph {et~al.}(2014)\citenamefont {Silva},
  \citenamefont {Monticone}, \citenamefont {Castaldi}, \citenamefont {Galdi},
  \citenamefont {Al{\`u}},\ and\ \citenamefont
  {Engheta}}]{silva2014performing}%
  \BibitemOpen
  \bibfield  {author} {\bibinfo {author} {\bibfnamefont {A.}~\bibnamefont
  {Silva}}, \bibinfo {author} {\bibfnamefont {F.}~\bibnamefont {Monticone}},
  \bibinfo {author} {\bibfnamefont {G.}~\bibnamefont {Castaldi}}, \bibinfo
  {author} {\bibfnamefont {V.}~\bibnamefont {Galdi}}, \bibinfo {author}
  {\bibfnamefont {A.}~\bibnamefont {Al{\`u}}},\ and\ \bibinfo {author}
  {\bibfnamefont {N.}~\bibnamefont {Engheta}},\ }\bibfield  {title} {\bibinfo
  {title} {Performing mathematical operations with metamaterials},\ }\href@noop
  {} {\bibfield  {journal} {\bibinfo  {journal} {Science}\ }\textbf {\bibinfo
  {volume} {343}},\ \bibinfo {pages} {160} (\bibinfo {year}
  {2014})}\BibitemShut {NoStop}%
\bibitem [{\citenamefont {Bykov}\ \emph {et~al.}(2014)\citenamefont {Bykov},
  \citenamefont {Doskolovich}, \citenamefont {Bezus},\ and\ \citenamefont
  {Soifer}}]{bykov2014optical}%
  \BibitemOpen
  \bibfield  {author} {\bibinfo {author} {\bibfnamefont {D.~A.}\ \bibnamefont
  {Bykov}}, \bibinfo {author} {\bibfnamefont {L.~L.}\ \bibnamefont
  {Doskolovich}}, \bibinfo {author} {\bibfnamefont {E.~A.}\ \bibnamefont
  {Bezus}},\ and\ \bibinfo {author} {\bibfnamefont {V.~A.}\ \bibnamefont
  {Soifer}},\ }\bibfield  {title} {\bibinfo {title} {Optical computation of the
  laplace operator using phase-shifted bragg grating},\ }\href@noop {}
  {\bibfield  {journal} {\bibinfo  {journal} {Optics Express}\ }\textbf
  {\bibinfo {volume} {22}},\ \bibinfo {pages} {25084} (\bibinfo {year}
  {2014})}\BibitemShut {NoStop}%
\bibitem [{\citenamefont {Ruan}(2015)}]{ruan2015spatial}%
  \BibitemOpen
  \bibfield  {author} {\bibinfo {author} {\bibfnamefont {Z.}~\bibnamefont
  {Ruan}},\ }\bibfield  {title} {\bibinfo {title} {Spatial mode control of
  surface plasmon polariton excitation with gain medium: from spatial
  differentiator to integrator},\ }\href@noop {} {\bibfield  {journal}
  {\bibinfo  {journal} {Optics Letters}\ }\textbf {\bibinfo {volume} {40}},\
  \bibinfo {pages} {601} (\bibinfo {year} {2015})}\BibitemShut {NoStop}%
\bibitem [{\citenamefont {Youssefi}\ \emph {et~al.}(2016)\citenamefont
  {Youssefi}, \citenamefont {Zangeneh-Nejad}, \citenamefont
  {Abdollahramezani},\ and\ \citenamefont {Khavasi}}]{youssefi2016analog}%
  \BibitemOpen
  \bibfield  {author} {\bibinfo {author} {\bibfnamefont {A.}~\bibnamefont
  {Youssefi}}, \bibinfo {author} {\bibfnamefont {F.}~\bibnamefont
  {Zangeneh-Nejad}}, \bibinfo {author} {\bibfnamefont {S.}~\bibnamefont
  {Abdollahramezani}},\ and\ \bibinfo {author} {\bibfnamefont {A.}~\bibnamefont
  {Khavasi}},\ }\bibfield  {title} {\bibinfo {title} {Analog computing by
  brewster effect},\ }\href@noop {} {\bibfield  {journal} {\bibinfo  {journal}
  {Optics Letters}\ }\textbf {\bibinfo {volume} {41}},\ \bibinfo {pages} {3467}
  (\bibinfo {year} {2016})}\BibitemShut {NoStop}%
\bibitem [{\citenamefont {Zhu}\ \emph {et~al.}(2017)\citenamefont {Zhu},
  \citenamefont {Zhou}, \citenamefont {Lou}, \citenamefont {Ye}, \citenamefont
  {Qiu}, \citenamefont {Ruan},\ and\ \citenamefont {Fan}}]{zhu2017plasmonic}%
  \BibitemOpen
  \bibfield  {author} {\bibinfo {author} {\bibfnamefont {T.}~\bibnamefont
  {Zhu}}, \bibinfo {author} {\bibfnamefont {Y.}~\bibnamefont {Zhou}}, \bibinfo
  {author} {\bibfnamefont {Y.}~\bibnamefont {Lou}}, \bibinfo {author}
  {\bibfnamefont {H.}~\bibnamefont {Ye}}, \bibinfo {author} {\bibfnamefont
  {M.}~\bibnamefont {Qiu}}, \bibinfo {author} {\bibfnamefont {Z.}~\bibnamefont
  {Ruan}},\ and\ \bibinfo {author} {\bibfnamefont {S.}~\bibnamefont {Fan}},\
  }\bibfield  {title} {\bibinfo {title} {Plasmonic computing of spatial
  differentiation},\ }\href@noop {} {\bibfield  {journal} {\bibinfo  {journal}
  {Nature Communications}\ }\textbf {\bibinfo {volume} {8}},\ \bibinfo {pages}
  {1} (\bibinfo {year} {2017})}\BibitemShut {NoStop}%
\bibitem [{\citenamefont {Zhang}\ \emph {et~al.}(2018)\citenamefont {Zhang},
  \citenamefont {Cheng}, \citenamefont {Wu}, \citenamefont {Wang},
  \citenamefont {Li},\ and\ \citenamefont {Zhang}}]{zhang2018implementing}%
  \BibitemOpen
  \bibfield  {author} {\bibinfo {author} {\bibfnamefont {W.}~\bibnamefont
  {Zhang}}, \bibinfo {author} {\bibfnamefont {K.}~\bibnamefont {Cheng}},
  \bibinfo {author} {\bibfnamefont {C.}~\bibnamefont {Wu}}, \bibinfo {author}
  {\bibfnamefont {Y.}~\bibnamefont {Wang}}, \bibinfo {author} {\bibfnamefont
  {H.}~\bibnamefont {Li}},\ and\ \bibinfo {author} {\bibfnamefont
  {X.}~\bibnamefont {Zhang}},\ }\bibfield  {title} {\bibinfo {title}
  {Implementing quantum search algorithm with metamaterials},\ }\href@noop {}
  {\bibfield  {journal} {\bibinfo  {journal} {Advanced Materials}\ }\textbf
  {\bibinfo {volume} {30}},\ \bibinfo {pages} {1703986} (\bibinfo {year}
  {2018})}\BibitemShut {NoStop}%
\bibitem [{\citenamefont {Guo}\ \emph {et~al.}(2018)\citenamefont {Guo},
  \citenamefont {Xiao}, \citenamefont {Minkov}, \citenamefont {Shi},\ and\
  \citenamefont {Fan}}]{guo2018photonic}%
  \BibitemOpen
  \bibfield  {author} {\bibinfo {author} {\bibfnamefont {C.}~\bibnamefont
  {Guo}}, \bibinfo {author} {\bibfnamefont {M.}~\bibnamefont {Xiao}}, \bibinfo
  {author} {\bibfnamefont {M.}~\bibnamefont {Minkov}}, \bibinfo {author}
  {\bibfnamefont {Y.}~\bibnamefont {Shi}},\ and\ \bibinfo {author}
  {\bibfnamefont {S.}~\bibnamefont {Fan}},\ }\bibfield  {title} {\bibinfo
  {title} {Photonic crystal slab laplace operator for image differentiation},\
  }\href@noop {} {\bibfield  {journal} {\bibinfo  {journal} {Optica}\ }\textbf
  {\bibinfo {volume} {5}},\ \bibinfo {pages} {251} (\bibinfo {year}
  {2018})}\BibitemShut {NoStop}%
\bibitem [{\citenamefont {Zhu}\ \emph {et~al.}(2019)\citenamefont {Zhu},
  \citenamefont {Lou}, \citenamefont {Zhou}, \citenamefont {Zhang},
  \citenamefont {Huang}, \citenamefont {Li}, \citenamefont {Luo}, \citenamefont
  {Wen}, \citenamefont {Zhu}, \citenamefont {Gong} \emph
  {et~al.}}]{zhu2019generalized}%
  \BibitemOpen
  \bibfield  {author} {\bibinfo {author} {\bibfnamefont {T.}~\bibnamefont
  {Zhu}}, \bibinfo {author} {\bibfnamefont {Y.}~\bibnamefont {Lou}}, \bibinfo
  {author} {\bibfnamefont {Y.}~\bibnamefont {Zhou}}, \bibinfo {author}
  {\bibfnamefont {J.}~\bibnamefont {Zhang}}, \bibinfo {author} {\bibfnamefont
  {J.}~\bibnamefont {Huang}}, \bibinfo {author} {\bibfnamefont
  {Y.}~\bibnamefont {Li}}, \bibinfo {author} {\bibfnamefont {H.}~\bibnamefont
  {Luo}}, \bibinfo {author} {\bibfnamefont {S.}~\bibnamefont {Wen}}, \bibinfo
  {author} {\bibfnamefont {S.}~\bibnamefont {Zhu}}, \bibinfo {author}
  {\bibfnamefont {Q.}~\bibnamefont {Gong}}, \emph {et~al.},\ }\bibfield
  {title} {\bibinfo {title} {Generalized spatial differentiation from the spin
  hall effect of light and its application in image processing of edge
  detection},\ }\href@noop {} {\bibfield  {journal} {\bibinfo  {journal}
  {Physical Review Applied}\ }\textbf {\bibinfo {volume} {11}},\ \bibinfo
  {pages} {034043} (\bibinfo {year} {2019})}\BibitemShut {NoStop}%
\bibitem [{\citenamefont {Zangeneh-Nejad}\ \emph {et~al.}(2020)\citenamefont
  {Zangeneh-Nejad}, \citenamefont {Sounas}, \citenamefont {Al{\`u}},\ and\
  \citenamefont {Fleury}}]{zangeneh2020analogue}%
  \BibitemOpen
  \bibfield  {author} {\bibinfo {author} {\bibfnamefont {F.}~\bibnamefont
  {Zangeneh-Nejad}}, \bibinfo {author} {\bibfnamefont {D.~L.}\ \bibnamefont
  {Sounas}}, \bibinfo {author} {\bibfnamefont {A.}~\bibnamefont {Al{\`u}}},\
  and\ \bibinfo {author} {\bibfnamefont {R.}~\bibnamefont {Fleury}},\
  }\bibfield  {title} {\bibinfo {title} {Analogue computing with
  metamaterials},\ }\href@noop {} {\bibfield  {journal} {\bibinfo  {journal}
  {Nature Reviews Materials}\ }\textbf {\bibinfo {volume} {6}},\ \bibinfo
  {pages} {207} (\bibinfo {year} {2020})}\BibitemShut {NoStop}%
\bibitem [{\citenamefont {Pierangeli}\ \emph
  {et~al.}(2020{\natexlab{a}})\citenamefont {Pierangeli}, \citenamefont
  {Marcucci}, \citenamefont {Brunner},\ and\ \citenamefont
  {Conti}}]{pierangeli2020noise}%
  \BibitemOpen
  \bibfield  {author} {\bibinfo {author} {\bibfnamefont {D.}~\bibnamefont
  {Pierangeli}}, \bibinfo {author} {\bibfnamefont {G.}~\bibnamefont
  {Marcucci}}, \bibinfo {author} {\bibfnamefont {D.}~\bibnamefont {Brunner}},\
  and\ \bibinfo {author} {\bibfnamefont {C.}~\bibnamefont {Conti}},\ }\bibfield
   {title} {\bibinfo {title} {Noise-enhanced spatial-photonic {Ising}
  machine},\ }\href@noop {} {\bibfield  {journal} {\bibinfo  {journal}
  {Nanophotonics}\ }\textbf {\bibinfo {volume} {9}},\ \bibinfo {pages} {4109}
  (\bibinfo {year} {2020}{\natexlab{a}})}\BibitemShut {NoStop}%
\bibitem [{\citenamefont {Pierangeli}\ \emph
  {et~al.}(2020{\natexlab{b}})\citenamefont {Pierangeli}, \citenamefont
  {Marcucci},\ and\ \citenamefont {Conti}}]{pierangeli2020adiabatic}%
  \BibitemOpen
  \bibfield  {author} {\bibinfo {author} {\bibfnamefont {D.}~\bibnamefont
  {Pierangeli}}, \bibinfo {author} {\bibfnamefont {G.}~\bibnamefont
  {Marcucci}},\ and\ \bibinfo {author} {\bibfnamefont {C.}~\bibnamefont
  {Conti}},\ }\bibfield  {title} {\bibinfo {title} {Adiabatic evolution on a
  spatial-photonic {Ising} machine},\ }\href@noop {} {\bibfield  {journal}
  {\bibinfo  {journal} {Optica}\ }\textbf {\bibinfo {volume} {7}},\ \bibinfo
  {pages} {1535} (\bibinfo {year} {2020}{\natexlab{b}})}\BibitemShut {NoStop}%
\bibitem [{\citenamefont {Pierangeli}\ \emph
  {et~al.}(2020{\natexlab{c}})\citenamefont {Pierangeli}, \citenamefont
  {Rafayelyan}, \citenamefont {Conti},\ and\ \citenamefont
  {Gigan}}]{pierangeli2020scalable}%
  \BibitemOpen
  \bibfield  {author} {\bibinfo {author} {\bibfnamefont {D.}~\bibnamefont
  {Pierangeli}}, \bibinfo {author} {\bibfnamefont {M.}~\bibnamefont
  {Rafayelyan}}, \bibinfo {author} {\bibfnamefont {C.}~\bibnamefont {Conti}},\
  and\ \bibinfo {author} {\bibfnamefont {S.}~\bibnamefont {Gigan}},\ }\bibfield
   {title} {\bibinfo {title} {Scalable spin-glass optical simulator},\
  }\href@noop {} {\bibfield  {journal} {\bibinfo  {journal} {arXiv preprint
  arXiv:2006.00828}\ } (\bibinfo {year} {2020}{\natexlab{c}})}\BibitemShut
  {NoStop}%
\bibitem [{\citenamefont {Kumar}\ \emph {et~al.}(2020)\citenamefont {Kumar},
  \citenamefont {Zhang},\ and\ \citenamefont {Huang}}]{kumar2020large}%
  \BibitemOpen
  \bibfield  {author} {\bibinfo {author} {\bibfnamefont {S.}~\bibnamefont
  {Kumar}}, \bibinfo {author} {\bibfnamefont {H.}~\bibnamefont {Zhang}},\ and\
  \bibinfo {author} {\bibfnamefont {Y.-P.}\ \bibnamefont {Huang}},\ }\bibfield
  {title} {\bibinfo {title} {Large-scale {Ising} emulation with four body
  interaction and all-to-all connections},\ }\href@noop {} {\bibfield
  {journal} {\bibinfo  {journal} {Communications Physics}\ }\textbf {\bibinfo
  {volume} {3}},\ \bibinfo {pages} {1} (\bibinfo {year} {2020})}\BibitemShut
  {NoStop}%
\bibitem [{\citenamefont {Nishimori}(2001)}]{nishimori2001statistical}%
  \BibitemOpen
  \bibfield  {author} {\bibinfo {author} {\bibfnamefont {H.}~\bibnamefont
  {Nishimori}},\ }\href@noop {} {\emph {\bibinfo {title} {Statistical physics
  of spin glasses and information processing: an introduction}}},\ \bibinfo
  {number} {111}\ (\bibinfo  {publisher} {Clarendon Press},\ \bibinfo {year}
  {2001})\BibitemShut {NoStop}%
\bibitem [{\citenamefont {Mertens}(1998)}]{mertens1998phase}%
  \BibitemOpen
  \bibfield  {author} {\bibinfo {author} {\bibfnamefont {S.}~\bibnamefont
  {Mertens}},\ }\bibfield  {title} {\bibinfo {title} {Phase transition in the
  number partitioning problem},\ }\href@noop {} {\bibfield  {journal} {\bibinfo
   {journal} {Physical Review Letters}\ }\textbf {\bibinfo {volume} {81}},\
  \bibinfo {pages} {4281} (\bibinfo {year} {1998})}\BibitemShut {NoStop}%
\bibitem [{\citenamefont {Wang}\ \emph {et~al.}(2013)\citenamefont {Wang},
  \citenamefont {Marandi}, \citenamefont {Wen}, \citenamefont {Byer},\ and\
  \citenamefont {Yamamoto}}]{wang2013coherent}%
  \BibitemOpen
  \bibfield  {author} {\bibinfo {author} {\bibfnamefont {Z.}~\bibnamefont
  {Wang}}, \bibinfo {author} {\bibfnamefont {A.}~\bibnamefont {Marandi}},
  \bibinfo {author} {\bibfnamefont {K.}~\bibnamefont {Wen}}, \bibinfo {author}
  {\bibfnamefont {R.~L.}\ \bibnamefont {Byer}},\ and\ \bibinfo {author}
  {\bibfnamefont {Y.}~\bibnamefont {Yamamoto}},\ }\bibfield  {title} {\bibinfo
  {title} {Coherent {Ising} machine based on degenerate optical parametric
  oscillators},\ }\href@noop {} {\bibfield  {journal} {\bibinfo  {journal}
  {Physical Review A}\ }\textbf {\bibinfo {volume} {88}},\ \bibinfo {pages}
  {063853} (\bibinfo {year} {2013})}\BibitemShut {NoStop}%
\bibitem [{\citenamefont {Strinati}\ \emph {et~al.}(2019)\citenamefont
  {Strinati}, \citenamefont {Bello}, \citenamefont {Pe'er},\ and\ \citenamefont
  {Dalla~Torre}}]{strinati2019theory}%
  \BibitemOpen
  \bibfield  {author} {\bibinfo {author} {\bibfnamefont {M.~C.}\ \bibnamefont
  {Strinati}}, \bibinfo {author} {\bibfnamefont {L.}~\bibnamefont {Bello}},
  \bibinfo {author} {\bibfnamefont {A.}~\bibnamefont {Pe'er}},\ and\ \bibinfo
  {author} {\bibfnamefont {E.~G.}\ \bibnamefont {Dalla~Torre}},\ }\bibfield
  {title} {\bibinfo {title} {Theory of coupled parametric oscillators beyond
  coupled {Ising} spins},\ }\href@noop {} {\bibfield  {journal} {\bibinfo
  {journal} {Physical Review A}\ }\textbf {\bibinfo {volume} {100}},\ \bibinfo
  {pages} {023835} (\bibinfo {year} {2019})}\BibitemShut {NoStop}%
\bibitem [{\citenamefont {Leuzzi}\ \emph {et~al.}(2009)\citenamefont {Leuzzi},
  \citenamefont {Conti}, \citenamefont {Folli}, \citenamefont {Angelani},\ and\
  \citenamefont {Ruocco}}]{leuzzi2009phase}%
  \BibitemOpen
  \bibfield  {author} {\bibinfo {author} {\bibfnamefont {L.}~\bibnamefont
  {Leuzzi}}, \bibinfo {author} {\bibfnamefont {C.}~\bibnamefont {Conti}},
  \bibinfo {author} {\bibfnamefont {V.}~\bibnamefont {Folli}}, \bibinfo
  {author} {\bibfnamefont {L.}~\bibnamefont {Angelani}},\ and\ \bibinfo
  {author} {\bibfnamefont {G.}~\bibnamefont {Ruocco}},\ }\bibfield  {title}
  {\bibinfo {title} {Phase diagram and complexity of mode-locked lasers: from
  order to disorder},\ }\href@noop {} {\bibfield  {journal} {\bibinfo
  {journal} {Physical Review Letters}\ }\textbf {\bibinfo {volume} {102}},\
  \bibinfo {pages} {083901} (\bibinfo {year} {2009})}\BibitemShut {NoStop}%
\bibitem [{\citenamefont {Bello}\ \emph {et~al.}(2019)\citenamefont {Bello},
  \citenamefont {Strinati}, \citenamefont {Dalla~Torre},\ and\ \citenamefont
  {Pe'er}}]{bello2019persistent}%
  \BibitemOpen
  \bibfield  {author} {\bibinfo {author} {\bibfnamefont {L.}~\bibnamefont
  {Bello}}, \bibinfo {author} {\bibfnamefont {M.~C.}\ \bibnamefont {Strinati}},
  \bibinfo {author} {\bibfnamefont {E.~G.}\ \bibnamefont {Dalla~Torre}},\ and\
  \bibinfo {author} {\bibfnamefont {A.}~\bibnamefont {Pe'er}},\ }\bibfield
  {title} {\bibinfo {title} {Persistent coherent beating in coupled parametric
  oscillators},\ }\href@noop {} {\bibfield  {journal} {\bibinfo  {journal}
  {Physical Review Letters}\ }\textbf {\bibinfo {volume} {123}},\ \bibinfo
  {pages} {083901} (\bibinfo {year} {2019})}\BibitemShut {NoStop}%
\bibitem [{\citenamefont {Hsueh}\ and\ \citenamefont
  {Sawchuk}(1978)}]{hsueh1978computer}%
  \BibitemOpen
  \bibfield  {author} {\bibinfo {author} {\bibfnamefont {C.}~\bibnamefont
  {Hsueh}}\ and\ \bibinfo {author} {\bibfnamefont {A.}~\bibnamefont
  {Sawchuk}},\ }\bibfield  {title} {\bibinfo {title} {Computer-generated
  double-phase holograms},\ }\href@noop {} {\bibfield  {journal} {\bibinfo
  {journal} {Applied Optics}\ }\textbf {\bibinfo {volume} {17}},\ \bibinfo
  {pages} {3874} (\bibinfo {year} {1978})}\BibitemShut {NoStop}%
\bibitem [{\citenamefont {Mendoza-Yero}\ \emph {et~al.}(2014)\citenamefont
  {Mendoza-Yero}, \citenamefont {M{\'\i}nguez-Vega},\ and\ \citenamefont
  {Lancis}}]{mendoza2014encoding}%
  \BibitemOpen
  \bibfield  {author} {\bibinfo {author} {\bibfnamefont {O.}~\bibnamefont
  {Mendoza-Yero}}, \bibinfo {author} {\bibfnamefont {G.}~\bibnamefont
  {M{\'\i}nguez-Vega}},\ and\ \bibinfo {author} {\bibfnamefont
  {J.}~\bibnamefont {Lancis}},\ }\bibfield  {title} {\bibinfo {title} {Encoding
  complex fields by using a phase-only optical element},\ }\href@noop {}
  {\bibfield  {journal} {\bibinfo  {journal} {Optics Letters}\ }\textbf
  {\bibinfo {volume} {39}},\ \bibinfo {pages} {1740} (\bibinfo {year}
  {2014})}\BibitemShut {NoStop}%
\bibitem [{\citenamefont {Ngcobo}\ \emph {et~al.}(2013)\citenamefont {Ngcobo},
  \citenamefont {Litvin}, \citenamefont {Burger},\ and\ \citenamefont
  {Forbes}}]{ngcobo2013digital}%
  \BibitemOpen
  \bibfield  {author} {\bibinfo {author} {\bibfnamefont {S.}~\bibnamefont
  {Ngcobo}}, \bibinfo {author} {\bibfnamefont {I.}~\bibnamefont {Litvin}},
  \bibinfo {author} {\bibfnamefont {L.}~\bibnamefont {Burger}},\ and\ \bibinfo
  {author} {\bibfnamefont {A.}~\bibnamefont {Forbes}},\ }\bibfield  {title}
  {\bibinfo {title} {A digital laser for on-demand laser modes},\ }\href@noop
  {} {\bibfield  {journal} {\bibinfo  {journal} {Nature Communications}\
  }\textbf {\bibinfo {volume} {4}},\ \bibinfo {pages} {1} (\bibinfo {year}
  {2013})}\BibitemShut {NoStop}%
\bibitem [{\citenamefont {Dudley}\ \emph {et~al.}(2012)\citenamefont {Dudley},
  \citenamefont {Vasilyeu}, \citenamefont {Belyi}, \citenamefont {Khilo},
  \citenamefont {Ropot},\ and\ \citenamefont {Forbes}}]{dudley2012controlling}%
  \BibitemOpen
  \bibfield  {author} {\bibinfo {author} {\bibfnamefont {A.}~\bibnamefont
  {Dudley}}, \bibinfo {author} {\bibfnamefont {R.}~\bibnamefont {Vasilyeu}},
  \bibinfo {author} {\bibfnamefont {V.}~\bibnamefont {Belyi}}, \bibinfo
  {author} {\bibfnamefont {N.}~\bibnamefont {Khilo}}, \bibinfo {author}
  {\bibfnamefont {P.}~\bibnamefont {Ropot}},\ and\ \bibinfo {author}
  {\bibfnamefont {A.}~\bibnamefont {Forbes}},\ }\bibfield  {title} {\bibinfo
  {title} {Controlling the evolution of nondiffracting speckle by complex
  amplitude modulation on a phase-only spatial light modulator},\ }\href@noop
  {} {\bibfield  {journal} {\bibinfo  {journal} {Optics Communications}\
  }\textbf {\bibinfo {volume} {285}},\ \bibinfo {pages} {5} (\bibinfo {year}
  {2012})}\BibitemShut {NoStop}%
\end{thebibliography}
%


\clearpage
\newpage

\setcounter{section}{0}

\newcommand{\hbAppendixPrefix}{S}
\renewcommand{\thefigure}{\hbAppendixPrefix\arabic{figure}}
\setcounter{figure}{0}

\renewcommand{\thetable}{\hbAppendixPrefix\arabic{table}}
\setcounter{table}{0}
\renewcommand{\theequation}{\hbAppendixPrefix\arabic{equation}}
\setcounter{equation}{0}

\onecolumngrid
\begin{center}
\large{\textbf{
Supplementary Material: Experimental Observation of Phase Transition in Spatial Photonic Ising Machine }}
\end{center}

%
%
%

\section{The experimental setup without gauge transformation, as presented in Fig. 1(b)}

The spin configuration $\mathbf{S}=\left\{\sigma_j \right\}$ is encoded on the SLM in an array of $N_x\times N_y$ macropixels. The size of a single macropixel is $\text{W}\times \text{W}$. As presented in Fig. 1(b), a paraxial beam with amplitude modulation $\{\xi_j\}$ illuminates on the SLM, which has the spatial phase modulation of $\{\sigma_j\}$. After reflected by SLM, the electric field is
\begin{equation}
E(\mathbf{x})=i\sum\limits_{j}{{{\xi }_{j}}{{\sigma }_{j}}}\text{rec}{{\text{t}}_{\text{W}}}(\mathbf{x}-{{\mathbf{x}}_{j}})=[i\sum\limits_{j}{{{\xi }_{j}}{{\sigma }_{j}}\delta (\mathbf{x}-{{\mathbf{x}}_{j}})}]\otimes \text{rec}{{\text{t}}_{\text{W}}}(\mathbf{x}) \label{eq:S1}.
\end{equation}
Here $\mathbf{x}=(x,y)$ denotes the spatial coordinate on the SLM plane, and ${{\mathbf{x}}_{j}}$ is the center position of the $j$th pixel on SLM, which takes the value of
\begin{center}
${{\mathbf{x}}_{j}}=\text{W} (m{{\mathbf{e}}_{x}}+n{{\mathbf{e}}_{y}})\text{ }\text{ }(1\le m\le {{N}_{x}},\text{ }1\le n\le {{N}_{y}})$.
\end{center}
In Eq. (\ref{eq:S1}), the notation $\otimes $ is the convolution operation, and the rectangular functions are defined as
\begin{center}
${\rm{rect}_{\rm{W}}}({\bf{x}}) = {\rm{rect}}(\frac{{\bf{x}}}{{\rm{W}}}) = \left\{
\begin{array}{l}
1\text{ }\text{ }\text{ }\left| x \right|,\left| y \right| \le {\rm{W}}/2\\
0\text{ }\text{ }\text{ }\left| x \right|,\left| y \right| > {\rm{W}}/2
\end{array} \right.$,
\end{center}
\begin{center}
${\text{rect}}({\mathbf{x}}) = \left\{ \begin{gathered}
  1\text{ }\text{ }\text{ }\left| x \right|,\left| y \right| \leqslant {\text{1}}/2 \hfill \\
  0\text{ }\text{ }\text{ }\left| x \right|,\left| y \right| > {\text{1}}/2 \hfill \\
\end{gathered}  \right.$.
\end{center}

According to the Fourier optics, the field at the back focal plane of a lens corresponds to the Fourier transform (FT) of $E(\mathbf{x})$,
\begin{center}
$E(\mathbf{x})\xrightarrow{\text{FT}}\tilde{E}(\mathbf{k})=\frac{1}{{{(2\pi )}^{2}}}\int_{-\infty }^{+\infty }{E(\mathbf{x}){{e}^{-i\mathbf{x}\cdot \mathbf{k}}}\text{d}\mathbf{x}}$.
\end{center}
Specifically,
\begin{center}
$\sum\limits_{j}{{{\xi }_{j}}{{\sigma }_{j}}\delta (\mathbf{x}-{{\mathbf{x}}_{j}})}\xrightarrow{\text{FT}}{{(\frac{1}{\text{W}})}^{2}}\sum\limits_{j}{{{\xi }_{j}}{{\sigma }_{j}}}{{e}^{i\mathbf{k}\cdot {{\mathbf{x}}_{j}}}}$
\end{center}
\begin{center}
$\text{rec}{{\text{t}}_{\text{W}}}(\mathbf{x})\xrightarrow{\text{FT}}{{\text{W}}^{2}}\text{sin}{{\text{c}}_{\text{W}}}(\mathbf{k})$
\end{center}
where
\begin{center}
$\text{sin}{{\text{c}}_{\text{W}}}(\mathbf{k})=\text{sinc}(\frac{\mathbf{k}\text{W}}{2\pi })=\frac{\sin \frac{\text{W}{{k}_{x}}}{2}}{\frac{W{{k}_{x}}}{2}}\cdot \frac{\sin \frac{W{{k}_{y}}}{2}}{\frac{W{{k}_{y}}}{2}}$, $\text{sinc}(\mathbf{k})=\frac{\sin \pi {{k}_{x}}}{\pi {{k}_{x}}}\frac{\sin \pi {{k}_{y}}}{\pi {{k}_{y}}}$.
\end{center}
Therefore we have
\begin{center}
$\tilde{E}(\mathbf{k})=(i\sum\limits_{j}{{{\xi }_{j}}{{\sigma }_{j}}}{{e}^{i\mathbf{k}\cdot {{\mathbf{x}}_{j}}}}\text{)}\cdot \text{sin}{{\text{c}}_{\text{W}}}(\mathbf{k})$.
\end{center}

The electric field on the detection plane, i.e. the back focal plane, is $E(u,v)$, where $\mathbf{u}=(u,v)$ is the spatial coordinate on the detection plane. Here the focal length of the FT lens is $f$ and the wavelength of the laser source is $\lambda $, and $u={{k}_{x}}\frac{f\lambda }{2\pi },\text{ }v={{k}_{y}}\frac{f\lambda }{2\pi }$. So
\begin{equation}
E(u,v)=(i\sum\limits_{j}{{{\xi }_{j}}{{\sigma }_{j}}}{{e}^{i\frac{2\pi }{f\lambda }\mathbf{u}\cdot {{\mathbf{x}}_{j}}}}\text{)}\cdot \text{sinc}(\frac{\mathbf{u}\text{W}}{f\lambda }).\label{eq:S2}
\end{equation}
The detected intensity image on CCD is
\begin{equation}
I(u,v)={{E}^{*}}(u,v)\cdot E(u,v)=\sum\limits_{ij}{{{\xi }_{i}}{{\xi }_{j}}{{\sigma }_{i}}}{{\sigma }_{j}}{{e}^{i\frac{2\pi }{f\lambda }({{\mathbf{x}}_{i}}-{{\mathbf{x}}_{j}})\cdot \mathbf{u}}}\text{sin}{{\text{c}}^{2}}(\frac{\mathbf{u}\text{W}}{f\lambda }). \label{eq:S3}
\end{equation}
The Hamiltonian is defined as
\begin{equation}
H=-J {I_0}=-J\cdot\sum\limits_{ij}{{{\xi }_{i}}{{\xi }_{j}}{{\sigma }_{i}}}{{\sigma }_{j}}, \label{eq:S4}
\end{equation}
where $J$ is a constant and ${{I}_{0}}$ is the intensity on the center of the detection plane, $(u,v)=(0,0)$. As a result, the optical spatial modulation method described as above successfully models the spin glass systems.

\section{Derivation of gauge transformation in Eq. (4)}

\begin{figure}[htb]
\centering
\includegraphics{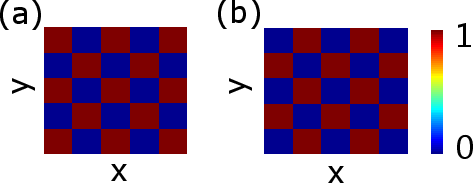}
\caption{ \label{fig:S2} Checkerboard patterns for functions (a) $M(\mathbf{x})\otimes \rm{rect}_{\text{W}}(\mathbf{x})$ and (b) $(1-M(\mathbf{x}))\otimes \rm{rec}\rm{t}_{\text{W}}(\mathbf{x})$.}
\end{figure}

Supposing that a collimated beam, with uniform amplitude, impinges on the phase-only SLM with the spatial phase modulation following Eq. (4) in the text, the beam wavefront becomes
\begin{equation}
\begin{aligned}
E({\bf{x}}) &= iM({\bf{x}})[\sum\limits_j {{\sigma _j}{e^{i{\alpha _j}}}\rm{rect}_{\rm{W}}({\bf{x}} - {{\bf{x}}_j})} ] + i(1 - M({\bf{x}}))[\sum\limits_j {{\sigma _j}{e^{ - i{\alpha _j}}}\rm{rect}_{\rm{W}}({\bf{x}} - {{\bf{x}}_j})} ]\\&= i[{H_1}({\bf{x}})M({\bf{x}}) + {H_2}({\bf{x}})(1 - M({\bf{x}}))] \otimes \rm{rect}_{\rm{W}}({\bf{x}})\label{eq:S5}
\end{aligned}
\end{equation}
where
\begin{center}
$\begin{array}{l}
{H_1}({\bf{x}}) = \sum\limits_j {{\sigma _j}{e^{i{{\cos }^{ - 1}}{\xi _j}}}\delta ({\bf{x}} - {{\bf{x}}_j})} \\
{H_2}({\bf{x}}) = \sum\limits_j {{\sigma _j}{e^{ - i{{\cos }^{ - 1}}{\xi _j}}}\delta ({\bf{x}} - {{\bf{x}}_j})}
\end{array}$
\end{center}
\begin{center}
${H_1}({\bf{x}}) + {H_2}({\bf{x}}) = 2\sum\limits_j {{\xi _j}{\sigma _j}\delta ({\bf{x}} - {{\bf{x}}_j})} $
\end{center}
\begin{center}
${H_1}({\bf{x}}) - {H_2}({\bf{x}}) = 2i\sum\limits_j {\sqrt {1 - \xi _j^2} {\sigma _j}\delta ({\bf{x}} - {{\bf{x}}_j})} $
\end{center}
and $M(\mathbf{x})$ and $1-M(\mathbf{x})$ are the checkerboard pattern functions as
\begin{center}
$\begin{array}{l}
M({\bf{x}}) = \sum\limits_{m,n =  - \infty }^{ + \infty } {[\delta (x - 2m{\rm{W}},y - 2n{\rm{W}})}  + \delta (x - (2m + 1){\rm{W}},y - (2n + 1){\rm{W}})]\\
1 - M({\bf{x}}) = \sum\limits_{m,n =  - \infty }^{ + \infty } {[\delta (x - 2m{\rm{W}},y - (2n + 1){\rm{W}}) + \delta (x - (2m + 1){\rm{W}},y - 2n{\rm{W}})} ]
\end{array}$.
\end{center}
The values of $M(\mathbf{x})\otimes \rm{rect}_{\text{W}}(\mathbf{x})$ and $(1-M(\mathbf{x}))\otimes \rm{rect}_{\text{W}}(\mathbf{x})$ are presented in Fig.~\ref{fig:S2}.
Specifically, in the case that all ${\xi _j}$s take binary values of either $+1$ or $-1$, the modulated optical field is simplified as
\begin{equation}
E({\bf{x}}) = i\sum\limits_j {{\sigma '_i}^z{\rm{rec}}{{\rm{t}}_{\rm{W}}}({\bf{x}} - {{\bf{x}}_j})}. \label{eq:S6}
\end{equation}

According to the Fourier optics, the lens L1(with focal length$f=100\text{mm}$) performs the spatial Fourier transformation(FT) on $E(\mathbf{x})$,
\begin{center}
$\begin{gathered}
  {H_1}({\mathbf{x}})\xrightarrow{{{\text{FT}}}}{h_1}({k_x},{k_y}) \hfill \\
  {H_2}({\mathbf{x}})\xrightarrow{{{\text{FT}}}}{h_2}({k_x},{k_y}) \hfill \\
\end{gathered} $
\end{center}
\begin{center}
$\begin{gathered}
  M({\mathbf{x}})\xrightarrow{{{\text{FT}}}}{(\frac{1}{{2{\text{W}}}})^2}\sum\limits_{m,n =  - \infty }^{ + \infty } {(1 + {e^{i{k_x}{\text{W}}}}{e^{i{k_y}{\text{W}}}})\delta ({k_x} - m\frac{\pi }{{\text{W}}},{k_y} - n\frac{\pi }{{\text{W}}})}  \hfill \\
  {\text{ }\text{ }\text{ }\text{ }\text{ }\text{ }\text{ }\text{ }\text{ }\text{ }\text{ }\text{ }\text{ }\text{ }} = {(\frac{1}{{2{\text{W}}}})^2}\sum\limits_{m,n =  - \infty }^{ + \infty } {(1 + {{( - 1)}^{m + n}})\delta ({k_x} - m\frac{\pi }{{\text{W}}},{k_y} - n\frac{\pi }{{\text{W}}})}  \hfill \\
\end{gathered} $
\end{center}
\begin{center}
$\begin{gathered}
  1 - M({\mathbf{x}})\xrightarrow{{{\text{FT}}}}{(\frac{1}{{2{\text{W}}}})^2}\sum\limits_{m,n =  - \infty }^{ + \infty } {({e^{i{k_x}{\text{W}}}} + {e^{i{k_y}{\text{W}}}})\delta ({k_x} - m\frac{\pi }{{\text{W}}},{k_y} - n\frac{\pi }{{\text{W}}})}  \hfill \\
  {\text{ }\text{ }\text{ }\text{ }\text{ }\text{ }\text{ }\text{ }\text{ }\text{ }\text{ }\text{ }\text{ }\text{ }} = {(\frac{1}{{2{\text{W}}}})^2}\sum\limits_{m,n =  - \infty }^{ + \infty } {({{( - 1)}^m} + {{( - 1)}^n})\delta ({k_x} - m\frac{\pi }{{\text{W}}},{k_y} - n\frac{\pi }{{\text{W}}})}  \hfill \\
\end{gathered} $.
\end{center}
So the spatial spectrum is
\begin{equation}
\begin{gathered}
  \tilde E({\mathbf{k}}) = \frac{i}{4}\{ {h_1}({k_x},{k_y}) \otimes [\sum\limits_{m,n =  - \infty }^{ + \infty } {(1 + {{( - 1)}^{m + n}})\delta ({k_x} - m\frac{\pi }{{\text{W}}},{k_y} - n\frac{\pi }{{\text{W}}})]}  \hfill \\
  {\text{ }\text{ }\text{ }\text{ }\text{ }\text{ }\text{ }\text{ }\text{ }\text{ }} + {h_2}({k_x},{k_y}) \otimes [\sum\limits_{m,n =  - \infty }^{ + \infty } {({{( - 1)}^m} + {{( - 1)}^n})\delta ({k_x} - m\frac{\pi }{{\text{W}}},{k_y} - n\frac{\pi }{{\text{W}}})]\} }  \hfill \\
  {\text{ }\text{ }\text{ }\text{ }\text{ }\text{ }\text{ }\text{ }\text{ }\text{ }} \cdot {\text{sin}}{{\text{c}}_{\text{W}}}({k_x},{k_y}) \hfill \\ \label{eq:S7}
\end{gathered} .
\end{equation}
The convolution terms state that the optical field on the focal plane is distributed periodically. This is because the pixels on the SLM have finite sizes, so $\tilde{E}(\mathbf{k})$ has the discrete Fourier transform (DFT)-like phenomena, where different diffraction orders are centered at $({{k}_{x}},{{k}_{y}})=(m,n)\frac{\pi }{\text{W}}$. The zeroth diffraction order is
\begin{equation}
\begin{aligned}
  {{\tilde E}_0}({\mathbf{k}}) &= \frac{i}{2}[{h_1}({k_x},{k_y}) + {h_2}({k_x},{k_y})] \cdot {\text{sin}}{{\text{c}}_{\text{W}}}({k_x},{k_y}) \\
   &= (\sum\limits_j {{\xi _j}{\sigma _j}} {e^{i{\mathbf{k}} \cdot {{\mathbf{x}}_j}}}{\text{)}} \cdot {\text{sin}}{{\text{c}}_{\text{W}}}({\mathbf{k}}) \\
   &= (\sum\limits_j {{\sigma '_j}^z} {e^{i{\mathbf{k}} \cdot {{\mathbf{x}}_j}}}{\text{)}} \cdot {\text{sin}}{{\text{c}}_{\text{W}}}({\mathbf{k}}) \\ \label{eq:S8}
\end{aligned}.
\end{equation}
In fact, the higher-diffraction-order fields may overlap with the zeroth-order field, thus the total field is the interference between the zeroth and higher diffraction orders,
\begin{equation}
\tilde E({\mathbf{k}}) = {\tilde E_0}({\mathbf{k}}) + \tilde E'({\mathbf{k}})\label{eq:S9}
\end{equation}
where $\tilde{{E}'}(\mathbf{k})$ is the contribution from higher diffraction orders. The field $\tilde{{E}'}(\mathbf{k})$ acts as noises that degrades the performance of the optical setup in modeling the spin glass systems.

The electric field on the detection plane, related to $\tilde E({\mathbf{k}})$ through $u = {k_x}\frac{{f\lambda }}{{2\pi }},{\text{ }}v = {k_y}\frac{{f\lambda }}{{2\pi }}$, is written as
\begin{equation}
{E_0}(u,v) = (\sum\limits_j {{\sigma '_j}^z} {e^{i\frac{{2\pi }}{{f\lambda }}{\mathbf{u}} \cdot {{\mathbf{x}}_j}}}{\text{)}} \cdot {\text{sinc}}(\frac{{{\mathbf{u}}{\text{W}}}}{{f\lambda }})\label{eq:S10}.
\end{equation}
If ${{H}_{1}}(\mathbf{x})$ and ${{H}_{2}}(\mathbf{x})$ are band-limited such that ${{E}_{0}}(u,v)$ is confined in the first Brillouin zone $(-\frac{f\lambda }{4\text{W}}\le u,v\le \frac{f\lambda }{4\text{W}})$, thus different diffraction orders do not overlap with each other. In the experiments, we are only interested in the detected field intensity within the finite area $(-\frac{f\lambda }{4\text{W}}\le u,v\le \frac{f\lambda }{4\text{W}})$, so only the zeroth diffraction order is contained and thus $E(u,v)={{E}_{0}}(u,v)$.

The detected intensity image on CCD is
\begin{equation}
I(u,v) = {E^*}(u,v) \cdot E(u,v) = \sum\limits_{ij} {{\sigma '_i}^z{\sigma '_j}^z} {e^{i\frac{{2\pi }}{{f\lambda }}({{\mathbf{x}}_i} - {{\mathbf{x}}_j}) \cdot {\mathbf{u}}}}{\text{sin}}{{\text{c}}^2}(\frac{{{\mathbf{u}}{\text{W}}}}{{f\lambda }})\label{eq:S11}.
\end{equation}
Then we arrive at the same result as Eq. \ref{eq:S3}. The Hamiltonian is defined as
\begin{equation}
H = -J\cdot{I_0} =  -J\cdot \sum\limits_{ij} {{\sigma '_i}^z{\sigma '_j}^z} \label{eq:S12}
\end{equation}
where $J$ is a constant with the unit of energy and ${{I}_{0}}$ is the intensity on the center of the detection plane $(u,v)=(0,0)$. As a result, the spatial spin glass systems are modeled with gauge transformation, where both encoding of the spin configuration and programming of the interaction strengths are realized by only one spatial phase modulator.

\section{Experimental setup and measurement of system Hamiltonian \label{sec:s3}}
\begin{figure}
\centering
\includegraphics{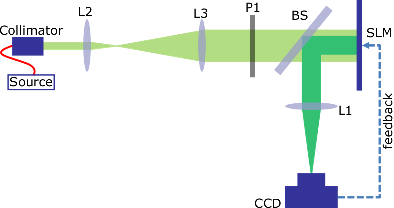}
\caption{ \label{fig:S1} Experimental setup of the spatial spin glass system with gauge transformation.}
\end{figure}

In the experiment, we use a green laser source (wavelength $\lambda =532\operatorname{nm}$) to generate a collimated beam with a planar wavefront and a uniform amplitude distribution (see the detailed setup in Fig. \ref{fig:S1}). Here the collimated laser with a beam waist radius of about 3.6mm is expanded by lenses L2(focal length is $50\text{mm}$) and L3(focal length is $500\text{mm}$), which generate a sufficiently wide Gaussian beam to cover the phase-only SLM (Holoeye PLUTO-NIR-011), such that the amplitude distribution at the used region of SLM is rather uniform. The polarizer P1 is used to prepare the incident beam linearly polarized along the long display axis of the SLM. A CCD (Ophir SP620) is used to detect the optical field intensity on the back focal plane.
We note that with a finite pixel-size detector, it is hard to exactly detect $I_0$ at the center, due to the restriction of the NA of lenses and the resolution of CCD. So the Hamiltonian is calculated by normalizing the detected intensity in a finite region as $H=-J{\iint\limits_{A}{[I(u,v)\times \bar{I}(u,v)]\text{d}u\text{d}v}}/{\iint\limits_{A}{\bar{I}(u,v)\text{d}u\text{d}v}}\;$, where $\bar{I}(u,v)$ is the field intensity on the detection plane when the SLM has uniform phase modulation of $\varphi_{j,\text{SLM}}=0$ in the squared detection region $A:|u|,|v|\le {d}/{2}\;$. We also estimate the impact of the noises by detecting a finite region on the detection plane instead of the center pixel in the experiments and find that the results are convergent and stable when $d>0.08\frac{f\lambda }{\text{W}}$. Here $f$ is the focal length of lens L1 shown in Fig.~\ref{fig:1}(d), and $\text{W}$ corresponds to the length of a macropixel on SLM encoding the effective spin configurations.

\section{Optical Metropolis-Hasting sampling \label{sec:s4}}
For the purpose of studying the statistical properties of spin glass systems, we first formulate the statistical ensembles containing sufficient samples of spin configurations $\mathbf{S}$. The statistical ensembles are obtained by optical iterations governed by Monte Carlo algorithm with Metropolis-Hasting sampling. In each iteration, a single spin is flipped and the updated spin configuration ${{\mathbf{S}}_{\text{new}}}$ is accepted with the probability of ${{e}^{-\Delta H/T}}$, depending on both the variance of the energy function $\Delta H=H({{\mathbf{S}}_{\text{new}}})-H(\mathbf{S})$ and the effective temperature as follow:
\begin{center}
 $\left\{ \begin{gathered}
  {\text{if }}\Delta H > 0:{\text{ accept }}{{\mathbf{S}}_{{\text{new}}}}{\text{ with the probability of }}{e^{ - \Delta H/T}} \hfill \\
  {\text{if }}\Delta H \leqslant 0:{\text{ accept }}{{\mathbf{S}}_{{\text{new}}}} \hfill \\
\end{gathered}  \right.$   .
\end{center}

\end{document}